\documentclass[a4paper,11pt]{article}

\usepackage[utf8]{inputenc}
\usepackage[final]{microtype}
\usepackage{lmodern}
\usepackage{exscale}
\usepackage[T1]{fontenc}

\usepackage{amsmath}
\usepackage{amssymb}
\usepackage{mathtools}

\usepackage[linktocpage]{hyperref}
\usepackage[dvipsnames,table]{xcolor}
\definecolor{CEICOorange}{RGB}{241,122,8}
\definecolor{CEICOblue}{RGB}{11,36,61}
\definecolor{CEICOcyan}{RGB}{12,183,209}
\hypersetup{
    colorlinks,
    linkcolor={CEICOorange},
    citecolor={CEICOorange},
    urlcolor={CEICOorange}
}

\usepackage[margin=2.5cm]{geometry}

\numberwithin{equation}{section}

\usepackage{graphicx}
\graphicspath{ {./figures/} }

\usepackage{cite}

\usepackage[labelfont=bf]{caption}

\usepackage{mathrsfs}

\usepackage{rotating}

\newcommand{\scr}{\scriptscriptstyle}

\usepackage{enumitem}

\usepackage{stmaryrd}

\usepackage{orcidlink}

\usepackage[low-sup]{subdepth}

\makeatletter
\newcommand{\dalembertian}{\mathop{\mathpalette\dalembertian@\relax}}
\newcommand{\dalembertian@}[2]{%
  \begingroup
  \sbox\z@{$\m@th#1\square$}%
  \dimen0=\fontdimen8
    \ifx#1\displaystyle\textfont\else
    \ifx#1\textstyle\textfont\else
    \ifx#1\scriptstyle\scriptfont\else
    \scriptscriptfont\fi\fi\fi3
  \makebox[\wd\z@]{%
    \hbox to \ht\z@{%
      \vrule width \dimen0
      \kern-\dimen0
      \vbox to \ht\z@{
        \hrule height \dimen0 width \ht\z@
        \vss
        \hrule height 2\dimen0
      }%
      \kern-2.5\dimen0
      \vrule width 2.5\dimen0
    }%
  }%
  \endgroup
}
\makeatother

\begin{document}

\begin{center}

\renewcommand{\thefootnote}{\fnsymbol{footnote}}

{\Large \bf 
On phase-space singular surfaces in $\boldsymbol{f(R)}$ gravity
}

\setcounter{footnote}{0} 

\bigskip

Dra\v{z}en Glavan\,\orcidlink{0000-0002-1983-0448}${}^{\,a,}$\footnote[1]{email: 
	\href{mailto:glavan@fzu.cz}{\tt glavan@fzu.cz}}
and David M.~J.~Vokrouhlick\'{y}\,\orcidlink{0009-0001-3648-8170}${}^{\,a,b,}$\footnote[2]{email: 
	\href{mailto:vokrouhlicky@fzu.cz}{\tt vokrouhlicky@fzu.cz}}

\bigskip

{\it 
${}^a$\,CEICO, Institute of Physics of the Czech Academy of Sciences (FZU),
\\
Na Slovance 1992/2, 182 21 Prague 8, Czech Republic

\medskip

${}^b$\,Institute of Theoretical Physics, Faculty of Mathematics and Physics, Charles University,\\
V Hole\v{s}ovi\v{c}k\'{a}ch 747/2, 180 00 Prague 8, Czech Republic
}

\

\

\parbox{0.88\linewidth}{
We perform a Hamiltonian constraint analysis of metric~$f(R)$ gravity in the Jordan frame and show that the regular constraint classification degenerates on singular phase-space surfaces located at~$f'(R)\!=\!0$ and~$f''(R)\!=\!0$. We then study the perturbative implications of these surfaces. For exact backgrounds satisfying~$f(R)\!=\!0$ and~$f'(R)\!=\!0$, the linearized spectrum is empty; the known pure~$R^2$ result is therefore a special case of a more general degeneracy in~$f(R)$ gravity. We also show that FLRW trajectories in the Starobinsky model can cross the surface~$f'(R)=0$, but that inhomogeneous perturbations develop a degenerate constraint structure at the crossing. The resulting crossing condition is better interpreted as a regularity condition for perturbative evolution than as an ordinary constraint within the Dirac--Bergmann algorithm. Together, these results distinguish backgrounds that lie entirely on a singular surface from backgrounds that cross one dynamically, and show that the two situations lead to different perturbative degeneracies.
}

\end{center}

\

\

\hrule

\tableofcontents

\vfill

\pagebreak

\section{Introduction}
\label{sec: Introduction}

The question of how many degrees of freedom propagate in a gravitational 
theory cannot in general be settled reliably by studying linear 
perturbations around a particular background. Around regular backgrounds 
such an analysis may reproduce the canonical count, but around 
degenerate backgrounds it can undercount the number of propagating 
modes. A definitive answer requires the full Hamiltonian constraint 
analysis, which determines the canonical structure of the theory
without relying on a background-perturbation split. This perspective 
proved essential in the case of pure~$R^2$ gravity, where the full 
theory propagates three degrees of freedom, while perturbations around 
Minkowski 
spacetime~\cite{Hell:2023mph,Golovnev:2023zen,Karananas:2024hoh} 
--- and, more generally, around traceless-Ricci backgrounds 
with~$R\!=\!0$~\cite{Barker:2025gon} --- exhibit an empty linearized 
spectrum due to a discontinuous change in the character of the 
constraints. In that case, perturbation theory fails precisely
because the relevant background lies on a singular surface in phase 
space, where the rank of the constraint system degenerates and the 
missing degrees of freedom become nonperturbative.

The aim of this paper is to extend this line of reasoning to
metric~$f(R)$ 
theories~\cite{Sotiriou:2008rp,DeFelice:2010aj,Nojiri:2010wj}
in the Jordan frame, and to identify the singular surfaces in phase 
space on which the Hamiltonian constraint structure changes 
discontinuously. These surfaces are characterized by a change in the 
rank of the constraint system, so that the canonical degree-of-freedom 
count cannot be inferred by continuity from nearby regular regions of 
phase space. They therefore signal a qualitative change in the dynamical 
system: constraints can change character, perturbative modes can become 
strongly coupled, and naive perturbative expansions can cease to be 
trustworthy. Our main interest is to determine these surfaces directly 
from the canonical structure of the theory, rather than infer
them indirectly from perturbative analyses around specially chosen 
backgrounds.

A crucial conceptual difference with respect to the pure~$R^2$ case
appears already at this stage. In pure~$R^2$ gravity, the singular
surface~$R \!=\! 0$ contains exact solutions of the equations of motion,
including Minkowski spacetime and more general traceless-Ricci
geometries. It is therefore possible to formulate perturbation theory
around such backgrounds, even though the resulting linearized theory is
singular, in the sense that it does not describe the dynamics in a
regular neighbourhood of phase space. By contrast, in a generic
metric~$f(R)$ theory the condition~$f'(R) \!=\! 0$ does not, by itself,
define a solution of the field equations, and it is therefore not
generally possible to treat such a configuration as a background for
perturbation theory.

Backgrounds satisfying~$f'(R)\!=\!0$ can solve the equations only if one
also has~$f(R)\!=\!0$. Constant-curvature backgrounds of this special
type were discussed in~\cite{Casado-Turrion:2023rni}, while the
corresponding linearized spectrum around maximally symmetric
representatives was also shown to be empty in~\cite{Casado-Turrion:2024esi}.
In general, however, the surface~$f'(R)\!=\!0$ is not a family of
backgrounds around which one can construct a conventional perturbative
spectrum. The issue is instead inherently dynamical: one must ask
whether physical phase-space trajectories can approach or cross such a
surface, and what this implies for the evolution of perturbations.

As we will show, this situation is not merely of formal interest. 
Although the singular surfaces identified by the Hamiltonian analysis 
are surfaces on which the constraint structure degenerates, they need 
not be dynamically inaccessible. In the cosmological system studied 
below, there exist background trajectories that pass smoothly through 
such surfaces. This raises a natural question: if the background 
evolution is regular while the perturbative description becomes
singular, how should one understand the behaviour of perturbations 
near the crossing? The Hamiltonian approach is particularly well suited 
to this problem, because it identifies the degeneracy directly and 
nonperturbatively, without requiring one to start from a specially 
chosen exact solution. In this respect the analysis also continues 
the logic of pure~$R^2$ gravity, where cosmological phase space contains 
trajectories that cross the singular surface~$R \!=\! 0$.

The paper is organized as follows. In 
Sec.~\ref{sec: Hamiltonian constraint analysis} we introduce the
canonical formulation (see~\cite{Deruelle:2009pu} and references 
therein for previous works on the subject) and carry out the 
full Dirac--Bergmann 
constraint analysis algorithm~\cite{Bergmann:1949,Dirac:1950pj,DiracBook}
for three cases of increasing generality: (i) the Starobinsky 
model, (ii) convex or concave~$f(R)$ models, and (iii) general~$f(R)$ 
models. In each case we identify the surfaces in phase space where the 
constraint structure changes discontinuously. Treating the three cases 
in parallel makes their common canonical structure manifest, while also 
isolating the degeneracies that arise only in the most general case.

In Sec.~\ref{sec: Perturbations around singular-surface solutions} we 
study a subclass of~$f(R)$ models that admit solutions satisfying
$f'(R)\!=\!0$, so that they lie entirely on the singular surface. We 
show that perturbations around these backgrounds have an empty 
linearized spectrum, through essentially the same mechanism as in 
pure~$R^2$ gravity~\cite{Barker:2025gon}. In
Sec.~\ref{sec: Perturbations near singular-surface crossings} we turn 
to the more subtle problem of perturbations around backgrounds that 
evolve through a singular surface. Focusing on the Starobinsky model, 
we show that its cosmological phase space contains such crossing 
trajectories, and we analyze the resulting breakdown of the perturbative 
constraint structure. We summarize our results and discuss their broader 
implications in Sec.~\ref{sec: Discussion}.

\section{Hamiltonian constraint analysis}
\label{sec: Hamiltonian constraint analysis}

In this section we consider three cases of~$f(R)$ models of
increasing generality. Such models are captured by the 
action
\begin{equation}
S[g_{\mu\nu}] = \int\! d^4x \, \sqrt{-g} \, f(R)
    \, ,
\label{IntroAction}
\end{equation}
with~$f$ an essentially arbitrary function. It is a dynamical theory of the metric~$g_{\mu\nu}$,
where~$g$ denotes its determinant, and~$R\!=\!g^{\mu\nu}R_{\mu\nu}$ is the Ricci scalar,
formed as the contraction of the Ricci tensor,~$R_{\mu\nu} \!=\! 2\partial_{[\rho} \Gamma^\rho_{\nu]\mu}
	\!+\! 2\Gamma_{\mu[\nu}^\rho \Gamma^{\sigma}_{\sigma]\rho}$, that itself is defined in terms
of the Christoffel symbols,~$\Gamma^\rho_{\mu\nu} \!=\! \frac{1}{2} g^{\rho\sigma}
	( \partial_\mu g_{\nu\sigma} \!+\! \partial_\nu g_{\mu\sigma} \!-\! \partial_\sigma g_{\mu\nu})$.
The equation of motion for such theories is
\begin{equation}
\Big( D_\mu D_\nu - g_{\mu\nu} D^\alpha D_\alpha \Big) f'(R)
    - R_{\mu\nu} f'(R)
    + \frac{1}{2} g_{\mu\nu} f(R)
    =
    0 \, ,
\label{CovariantEOM}
\end{equation}
where~$D_\mu$ is a covariant derivative with respect to the metric~$g_{\mu\nu}$.

After briefly recalling the Arnowitt--Deser--Misner decomposition
we construct the Jordan-frame Hamiltonian
formulation of the three classes of~$f(R)$ theories:
the Starobinsky model, convex or concave~$f(R)$ models, and
completely general~$f(R)$ models, 
and perform the corresponding Dirac--Bergmann
constraint analysis. The analysis reveals singular surfaces,
located at~$f'(R)\!=\!0$ and~$f''(R)\!=\!0$, on which the Hamiltonian
constraint structure changes discontinuously. These surfaces also
coincide with singular points of the transformation relating the
Jordan and Einstein frames. We emphasize, however, that the change
in constraint character found below is not a consequence of this
frame transformation, but is instead an intrinsic property of the
Jordan-frame canonical theory. Furthermore, the two frames are known 
not to be equivalent globally~\cite{Bahamonde:2016wmz,Alho:2016gzi,Rinaldi:2018qpu}.
It has also been pointed out that this issue can influence the analysis of 
linearized perturbations~\cite{Hell:2025lbl}.

Although the Dirac--Bergmann algorithm is conceptually
straightforward, its implementation can be technically cumbersome,
and the simplicity of the analysis depends strongly on the choice
of variables. In fact, each of the three models considered below
is most naturally formulated in a different way, with a different
number of auxiliary fields. It is therefore useful to present the
three cases separately, even though the completely general~$f(R)$
theory formally contains the preceding two as special cases.
The separate treatments make the constraint structure more 
transparent, and show explicitly how the simpler models are 
embedded in the general analysis.

\subsection{ADM decomposition}
\label{sec: ADM decomposition}

We begin with the Arnowitt--Deser--Misner (ADM)
decomposition~\cite{Arnowitt:1962hi} of the metric. The lapse
function, shift vector, and spatial metric are introduced by
identifying the following metric components:
\begin{equation}
g^{00} = - \frac{1}{N^2} \, ,
\qquad \qquad
g_{0i} = N_i \, ,
\qquad \qquad
g_{ij} = h_{ij} \, ,
\end{equation}
which in turn determines the remaining components as
\begin{equation}
g_{00} = - N^2 + N_i N^i \, ,
\qquad \qquad
g^{0i} = \frac{N^i}{N^2} \, ,
\qquad \qquad
g^{ij} = h^{ij} - \frac{N^i N^j}{N^2} \, .
\end{equation}

The first time derivative of the spatial metric is encoded in the
extrinsic curvature,
\begin{equation}
K_{ij} = - \frac{1}{2N} \Bigl( \dot{h}_{ij} - \nabla_i N_j - \nabla_j N_i \Bigr) \, ,
\label{Kdef}
\end{equation}
where~$\nabla_i$ stands for a covariant derivative on the spatial
slice with an induced metric~$h_{ij}$ and the induced Christoffel
symbol~$\gamma_{ij}^k \!=\! \frac{1}{2} h^{k\ell} ( \partial_i h_{j\ell}
    \!+\! \partial_j h_{i\ell} \!-\! \partial_\ell h_{ij})$.
The second time derivative is conveniently captured by the variable
introduced in~\cite{Buchbinder:1987vp},
\begin{equation}
F_{i j} = - \frac{1}{N} \dot{K}_{ij}
	- K_{ik} {K^k}_j
	+ \frac{N^k}{N} \nabla_k K_{ij}
	+ \frac{2}{N} K_{k(i} \nabla_{j)} N^k
	- \frac{1}{N} \nabla_i \nabla_j N \, .
\label{Fdef}
\end{equation}
With these definitions, the four-dimensional Ricci scalar
decomposes as
\begin{equation}
R = 2F + K^2 - K^{ij} K_{ij} + \mathcal{R} \, ,
\label{Rdecomposition}
\end{equation}
where the Ricci scalar on a spatial 
slice,~$\mathcal{R} \!=\! h^{ij} \mathcal{R}_{ij}$, is a contraction
of the Ricci tensor,~$\mathcal{R}_{ij} \!=\! \partial_k \gamma^k_{ij}
    \!-\! \partial_j \gamma^k_{ik}
    \!+\! \gamma^k_{ij} \gamma^{\ell}_{k\ell}
    \!-\! \gamma^k_{i\ell} \gamma^{\ell}_{kj}$.

\subsection{Starobinsky model}
\label{subsec: Starobinsky model}

We first consider the special case of~$f(R)$ gravity given by the
Starobinsky model~\cite{Starobinsky:1979ty},
\begin{equation}
S[g_{\mu\nu}] = \int\! d^4x \, \sqrt{-g} \,
	\biggl[ \frac{R}{\kappa^2} + \alpha R^2 \biggr]
	\, ,
\label{StarobinskyAction}
\end{equation}
where we take~$\kappa^2\!>\!0$, while~$\alpha$ is allowed to have
either sign, excluding only the case~$\alpha\!=\!0$. The latter
corresponds to the pure Einstein-Hilbert action, which is not
smoothly connected to the theory above from the point of view of
the number of propagating degrees of freedom.

This model provides the simplest setting, beyond the pure~$R^2$
theory considered in~\cite{Barker:2025gon}, in which the general
questions posed in this paper can be studied explicitly. In
particular, it already exhibits singular surfaces in phase space
on which the Hamiltonian constraint structure changes
discontinuously, while still being simple enough to allow for a
direct analysis of the corresponding cosmological dynamics. For
this reason, it serves as a natural starting point before turning
to more general classes of~$f(R)$ theories.

\subsubsection{Canonical action}
\label{subsubsec: Canonical action}

We now construct the canonical form of the theory
following the recipe from~\cite{Gitman}. Starting from
the ADM decomposition of the action in~(\ref{StarobinskyAction}),
\begin{equation}
S \bigl[ N, N_i, h_{ij} \bigr]
	= 
	\int\! d^4 x \, N \sqrt{h} \, 
	\biggl[
	\frac{1}{\kappa^2}
	\Bigl( 2F \!+\! K^2 \!-\! K^{ij} K_{ij} \!+\! \mathcal{R} \Bigr)
	+
	\alpha
	\Bigl( 2F \!+\! K^2 \!-\! K^{ij} K_{ij} \!+\! \mathcal{R} \Bigr)^{\!2}
	\biggr]
	\, ,
\label{ADMactionStar}
\end{equation}
we promote the time-derivative variables to independent velocity
fields,
\begin{equation}
K_{ij} \longrightarrow \mathcal{K}_{ij} \, ,
\qquad \quad
F_{ij} \longrightarrow \mathcal{F}_{ij} \, ,
\end{equation}
and introduce the Lagrange multipliers~$\pi^{ij}$ and~$\rho^{ij}$.
These enforce the on-shell equivalence of the extended formulation
with the original higher-derivative theory. The extended action
then reads
\begin{align}
\MoveEqLeft[3]
\mathcal{S} \big[ N, N_i, h_{ij}, \mathcal{K}_{ij}, \pi^{ij} , \mathcal{F}_{ij}, \rho^{ij} \big] 
    =
    \int\! d^4x \, \bigg\{
    N \sqrt{h} \, 
	\bigg[
	\frac{1}{\kappa^2}
	\Bigl( 2\mathcal{F} \!+\! \mathcal{K}^2 \!-\! \mathcal{K}^{ij} \mathcal{K}_{ij} \!+\! \mathcal{R} \Bigr)
\nonumber \\
&
	+
	\alpha
	\Bigl( 2\mathcal{F} \!+\! \mathcal{K}^2 \!-\! \mathcal{K}^{ij} \mathcal{K}_{ij} \!+\! \mathcal{R} \Bigr)^{\!2}
	\bigg]
    +
    \pi^{ij} \Big( \dot{h}_{ij} + 2N \mathcal{K}_{ij} - 2 \nabla_{(i} N_{j)} \Big)
\nonumber \\
&
    +
    \rho^{ij} \Big(
    \dot{\mathcal{K}}_{ij}
    +
    N \mathcal{K}_{ik} \mathcal{K}^k{}_j
    -
    N^k \nabla_k \mathcal{K}_{ij}
    -
    2 \mathcal{K}_{k(i} \nabla_{j)} N^k
    +
    \nabla_i \nabla_j N
    +
    N \mathcal{F}_{ij}
    \Big)
    \bigg\}
    \, .
\label{StarExtendedAction}
\end{align}

The canonical action is obtained by solving algebraically for the
velocity fields wherever possible. In the present case,
\(\mathcal{F}_{ij}\) enters the action only through its trace,
apart from the term~$\rho^{ij}\mathcal{F}_{ij}$. Hence
the equation obtained by varying with respect to~$\mathcal{F}_{ij}$
determines only the trace of~$\mathcal{F}_{ij}$:
\begin{align}
\MoveEqLeft[4]
\frac{\delta \mathcal{S}}{\delta \mathcal{F}_{ij}}
    =
    N \sqrt{h} \bigg[ \frac{2 h^{ij}}{\kappa^2}
    +
    4\alpha h^{ij}
    \Bigl( 2\mathcal{F} \!+\! \mathcal{K}^2 \!-\! \mathcal{K}^{ij} \mathcal{K}_{ij} \!+\! \mathcal{R} \Bigr)
    +
    \frac{\rho^{ij}}{\sqrt{h}}
    \bigg]
    \approx 0 \, ,
\\
&
    \Longrightarrow
    \qquad
    \mathcal{F}
    \approx
    \overline{\mathcal{F}}
    =
    -
    \frac{1}{24\alpha} \Big( \frac{\rho}{\sqrt{h}} + \frac{6}{\kappa^2} \Big)
    -
    \frac{1}{2} \big( \mathcal{K}^2 - \mathcal{K}^{ij} \mathcal{K}_{ij} + \mathcal{R} \big)
    \, .
\end{align}
Substituting this solution for~$\mathcal{F}$ back into the
extended action~(\ref{StarExtendedAction}) as a strong equality,
we obtain the canonical action
\begin{align}
\MoveEqLeft[5]
\mathscr{S} \big[ N, N_i, \lambda_{ij}, h_{ij} , \pi^{ij}, \mathcal{K}_{ij} , \rho^{ij} \big] 
    \equiv\mathcal{S} \big[ N, N_i, h_{ij}, \mathcal{K}_{ij}, \pi^{ij} , 
        \mathcal{F}_{ij} \!\to\! \tfrac{1}{3} h_{ij} \overline{\mathcal{F}} \!-\! \lambda_{ij}, \rho^{ij}\big]
\nonumber \\
    ={}&
    \int\! d^4x \, \Big[
    \pi^{ij} \dot{h}_{ij}
    +
    \rho^{ij} \dot{\mathcal{K}}_{ij}
    -
    N \big( \mathcal{H} + \lambda_{ij} \Phi^{ij} \big)
    -
    N_{i} \mathcal{H}^i
    \Big]
    \, ,
\label{StarCanonicalAction}
\end{align}
where the traceless part of~$\mathcal{F}_{ij}$ has been relabeled
as~$(-\lambda_{ij})$ in order to make explicit its role as the
Lagrange multiplier associated with the primary traceless
constraint,
\begin{equation}
\Phi^{ij} = H^{ij}{}_{k\ell} \rho^{k\ell} \, ,
\qquad \quad
H^{ij}{}_{k\ell}
    =
    \delta^{(i}_k \delta^{j)}_\ell
    -
    \frac{1}{3} h^{ij} h_{k\ell}
    \, .
\label{StarPrimaryTraceless}
\end{equation}
Note that we work in the reduced phase space in which lapse, shift, 
and the remaining multiplier fields are treated as Lagrange multipliers 
rather than canonical variables. In this reduced formulation, we refer 
to the constraints appearing directly in the canonical action as the 
initial constraints.

After using the freedom to shift~$\lambda_{ij}$ so as to eliminate
the traceless part of~$\rho^{ij}$ from them, the tentative
Hamiltonian and momentum constraints are
\begin{align}
\mathcal{H} ={}&
    \sqrt{h}
    \bigg[
    -
    2 \mathcal{K}_{ij} \frac{\pi^{ij}}{\sqrt{h}}
    +
    \frac{1}{144\alpha}
	\Big( \frac{6}{\kappa^2} + \frac{\rho}{\sqrt{h}} \Big)^{\!2}
    -
    \frac{1}{6}
    \Big( 2 \nabla^2 + 3 \mathcal{K}^{ij} \mathcal{K}_{ij} - \mathcal{K}^2 - \mathcal{R} \Big)
    \frac{\rho}{\sqrt{h}}
    \bigg]
    \, ,
\\
\mathcal{H}^i ={}&
    \sqrt{h}
    \bigg[
    -
    2 \nabla_{j} \Big( \frac{\pi^{ij}}{\sqrt{h}} \Big)
    +
    \frac{ \nabla^i \mathcal{K} }{3} \frac{\rho}{\sqrt{h}}
    -
    \frac{2}{3} \nabla_j \Big( \mathcal{K}^{ij} \frac{\rho}{\sqrt{h}} \Big)
    \bigg]
    \, .
\end{align}
This canonical formulation is the starting point for the full
constraint analysis, which we now carry out using the
Dirac--Bergmann algorithm.

\subsubsection{Constraint analysis}

We now apply the Dirac--Bergmann algorithm to the canonical
formulation derived in the previous subsection. The canonical phase
space is spanned by the pairs~$\bigl(h_{ij},\pi^{ij}\bigr)$
and~$\bigl(\mathcal{K}_{ij},\rho^{ij}\bigr)$, with canonical
Poisson brackets
\begin{equation}
\big\{ h_{i j}(t, \vec{x}\,) ,
    \pi^{k \ell} (t, \vec{x}^{\,\prime} ) \big\}
    =
    \delta_{(i}^k \delta_{j)}^\ell 
        \delta^3 (\vec{x} \!-\! \vec{x}^{\,\prime} )
    \, ,
\qquad
\big\{ \mathcal{K}_{i j}(t, \vec{x} \,), 
    \rho^{k \ell} (t, \vec{x}^{\,\prime} ) \big\}
    =
    \delta_{(i}^k \delta_{j)}^\ell
    \delta^3(\vec{x} \!-\! \vec{x}^{\,\prime} )
    \, .
\end{equation}
To streamline the analysis, we work with smeared constraints. For
arbitrary test functions~$q$,~$q_i$, and traceless~$q_{ij}$, we
define
\begin{equation}
\mathcal{H} \llbracket q \rrbracket \equiv\!
    \int\! d^3 x \, q(x) \mathcal{H}(x) \, , 
\qquad
    \mathcal{H}^i \llbracket q_i \rrbracket \equiv\! 
        \int\! d^3 x \, q_i(x) \mathcal{H}^i(x) \, , 
\qquad 
    \Phi^{i j} \llbracket q_{i j} \rrbracket \equiv\!
        \int\! d^3 x \, q_{i j}(x) \Phi^{i j}(x) \, .
\label{SmearingDef}
\end{equation}
In this notation, the total Hamiltonian is
\begin{equation}
H_{\rm tot} = \mathcal{H} \llbracket N \rrbracket 
    + \mathcal{H}^i \llbracket N_i \rrbracket 
    + \Phi^{ij} \llbracket N\lambda_{ij} \rrbracket \, ,
\end{equation}
and generates time evolution through Hamilton's equations. The
first step in the Dirac--Bergmann procedure is therefore to
evaluate the Poisson brackets among the primary constraints. These
brackets are summarized in Table~\ref{PrimaryBracketsTable1},
\begin{table}[h!]
\renewcommand{\arraystretch}{2}
\setlength\tabcolsep{0pt}
\centering
\begin{tabular}{ | w{c}{1.5cm} || w{c}{5.5cm} w{c}{5.5cm} w{c}{3.4cm} |} 
\hline
	&
	$\mathcal{H} \llbracket s \rrbracket$
	&
	$\mathcal{H}^k \llbracket s_k \rrbracket$
	&
	$\Phi^{k\ell} \llbracket s_{k\ell} \rrbracket $
\\
\hline\hline
	$\mathcal{H} \llbracket q \rrbracket $
	&
	$2\Psi^{ij} \big\llbracket s\nabla\!_i \nabla\!_j q 
        \!-\! q\nabla\!_i\nabla\!_j s \big\rrbracket $
	&
	$2 \Psi^{ij} \bigl\llbracket 
        2q \mathcal{K}_{i}^{k} \nabla_j s_k \!+\! q s_k \nabla^k \mathcal{K}_{ij} \bigr\rrbracket$
	&
	$-2 \Psi^{ij} \llbracket s_{ij}q \rrbracket $
\\
	$\mathcal{H}^i \llbracket q_i \rrbracket$
	&
	$- 2 \Psi^{ij} \bigl\llbracket 2s \mathcal{K}_{i}{}^{k}
        \nabla_j q_k \!+\! s q_k \nabla^k \mathcal{K}_{ij} \bigr\rrbracket$
	&
	$0$
	&
	$0$
\\
	$\Phi^{ij} \llbracket q_{ij} \rrbracket $
	&
	$2 \Psi^{ij} \llbracket q_{ij}s \rrbracket $
	&
	$0$
	&
	$0$
\\[0.8ex]
\hline
\end{tabular}
\caption{Poisson brackets among the primary constraints in the
Starobinsky model. Each entry gives~$\tt \{ row , column \} \approx entry$.}
\label{PrimaryBracketsTable1}
\end{table}
where we have introduced
\begin{equation}
\Psi^{ij} = H^{ij}{}_{k\ell} \Big( \pi^{k\ell}
        + \frac{\rho}{6} \mathcal{K}^{k\ell} \Big) \, .
\end{equation}
Table~\ref{PrimaryBracketsTable1} shows that conservation of the
primary constraints requires this quantity to vanish weakly. 
It therefore appears as a secondary constraint,
\begin{equation}
\Psi^{ij} \approx 0 \, .
\end{equation}
Note that henceforth we employ Dirac's notation for strong (off~shell)
and weak (on-shell) equalities~\cite{DiracBook}.

We must now check whether conservation of the secondary constraint
generates further constraints or instead determines the Lagrange
multiplier~$\lambda_{ij}$. Its brackets with the primary
constraints are
\begin{subequations}
\begin{align}
\big\{ \Psi^{ij} \llbracket q_{ij} \rrbracket ,
    \mathcal{H} \llbracket s \rrbracket \big\} \approx{}&
    -
    \frac{1}{3} \pi \big\llbracket H^{ijk\ell} \mathcal{K}_{k\ell} q_{ij} s \big\rrbracket
    -
    \frac{1}{9} \rho \big\llbracket
        H^{ijk\ell} \big( 3 \mathcal{K}_{km} \mathcal{K}^m{}_\ell
            \!-\! \mathcal{K}_{k\ell} \mathcal{K} \big) q_{ij} s
        \big\rrbracket
\nonumber \\
&
    +
    \frac{1}{6} 
    \rho \big\llbracket
    H^{ijk\ell} s (\mathcal{R}_{k\ell} - \nabla_k \nabla_\ell ) q_{ij}
    \big\rrbracket
    -
    \frac{1}{3} \rho\big\llbracket H^{ijk\ell} \nabla_k \big( q_{ij} \nabla_\ell s \big) \big\rrbracket
    \, ,
\\
\big\{ \Psi^{ij} \llbracket q_{ij} \rrbracket ,
    \mathcal{H}^k \llbracket s_k \rrbracket \big\} 
    \approx{}&
    \frac{1}{3} \rho \big\llbracket H^{ijk\ell} q_{ij} \mathcal{K}_{k}{}^m \nabla_\ell s_m \big\rrbracket
    +
    \frac{1}{6} \rho \big\llbracket H^{ijk\ell} q_{ij} s_m \nabla^m \mathcal{K}_{k\ell} \big\rrbracket
    \, ,
\\
\big\{ \Psi^{ij} \llbracket q_{ij} \rrbracket ,
    \Phi^{k\ell} \llbracket s_{k\ell} \rrbracket \big\}
    \approx{}&
    - \frac{1}{6} \rho \big\llbracket H^{ijk\ell} q_{ij} s_{k\ell} \big\rrbracket
    \, ,
\\
\big\{ \Psi^{ij} \llbracket q_{ij} \rrbracket ,
    \Psi^{k\ell} \llbracket s_{k\ell} \rrbracket \big\} \approx{}&
    0 \, .
\end{align}
\label{StarAlgebraSecondary}%
\end{subequations}
Rather than generating a tertiary constraint, the conservation of~$\Psi^{ij}$
determines the traceless Lagrange multiplier~$\lambda_{ij}$ on shell,
\begin{align}
\lambda_{ij} \approx \overline{\lambda}_{ij} ={}&
    H_{ij}{}^{k\ell} \bigg[
    \mathcal{R}_{k\ell} - 2  \mathcal{K}_{km} \mathcal{K}^m{}_\ell
            + \frac{2}{3} \mathcal{K}_{k\ell} \mathcal{K}
            - \frac{ \nabla_k \nabla_\ell N }{N}
    -
    \frac{2 \pi \mathcal{K}_{k\ell}}{\rho}
\nonumber \\
&   \hspace{3cm}
    -
    \frac{\sqrt{h}}{\rho} \nabla_k \nabla_\ell \Big( \frac{\rho}{\sqrt{h}} \Big)
    +
    \frac{ 2 \mathcal{K}_{k}{}^m \nabla_\ell N_m
    +
    N_m \nabla^m \mathcal{K}_{k\ell} }{N} 
    \bigg]
    \, .
\label{StarLambdaSolution}
\end{align}
This exhausts the Dirac--Bergmann algorithm: no further generations
of constraints arise.

The final step is to identify the first-class combinations of
constraints explicitly. Although this is not necessary for closing
the algorithm, it is useful because it makes the constraint
structure more transparent. We do this by shifting the traceless
Lagrange multiplier by its on-shell value,
\begin{equation}
\lambda_{ij} \longrightarrow \lambda_{ij} + \overline{\lambda}_{ij} \, ,
\end{equation}
which induces the following redefinition of the Hamiltonian and
momentum constraints:
\begin{align}
&
\boldsymbol{\mathcal{H}} = 
    \mathcal{H}
    +
    \bigg[
    \mathcal{R}_{ij} - 2  \mathcal{K}_{ik} \mathcal{K}^k{}_j
            + \frac{2}{3} \mathcal{K}_{ij} \mathcal{K}
            - \nabla_i \nabla_j
    -
    \frac{2 \pi \mathcal{K}_{ij} }{\rho}
    -
    \frac{\sqrt{h}}{\rho} \nabla_i \nabla_j \Big( \frac{\rho}{\sqrt{h}} \Big)
    \bigg]
    \Phi^{ij}
    \, ,
\\
&
\boldsymbol{\mathcal{H}}^i =
    \mathcal{H}^i
    +
    \Phi^{jk} \nabla^i \mathcal{K}_{jk}
    -
    2 \sqrt{h} \nabla_j \Big( \mathcal{K}^i{}_{k} \frac{ \Phi^{jk} }{ \sqrt{h} } \Big)
    \, ,
\end{align}
for which the first-class character becomes manifest. The final canonical
constraint algebra is summarized in Table~\ref{FinalBrackets1}.
\begin{table}[h!]
\renewcommand{\arraystretch}{2}
\setlength\tabcolsep{0pt}
\centering
\begin{tabular}{ | w{c}{1.5cm} || w{c}{3.5cm} w{c}{3.5cm} w{c}{3.5cm} w{c}{3.5cm} |} 
\hline
	&
	$\boldsymbol{\mathcal{H}} \llbracket s \rrbracket $
	&
	$\boldsymbol{\mathcal{H}}^k \llbracket s_k \rrbracket$
	&
	$\Phi^{k\ell} \llbracket s_{k\ell} \rrbracket$
	&
	$\Psi^{k\ell} \llbracket s_{k\ell} \rrbracket$
\\
\hline\hline
	$\boldsymbol{\mathcal{H}} \llbracket q \rrbracket$
	&
	$0$
	&
	$0$
	&
	$0$
	&
	$0$
\\
	$\boldsymbol{\mathcal{H}}^i \llbracket q_i \rrbracket$
	&
	$0$
	&
	$0$
	&
	$0$
	&
	$0$
\\
	$\Phi^{ij} \llbracket q_{ij} \rrbracket$
	&
	$0$
	&
	$0$
	&
    \cellcolor{black!15}
	$0$
    &
    \cellcolor{black!15}
    $\frac{1}{6} \rho \big\llbracket H^{ijk\ell} q_{ij} s_{k\ell} \big\rrbracket$
\\
	$\Psi^{ij} \llbracket q_{ij} \rrbracket $
	&
	$0$
	&
	$0$
	&
    \cellcolor{black!15}
	$- \frac{1}{6} \rho \big\llbracket H^{ijk\ell} q_{ij} s_{k\ell} \big\rrbracket $
    &
    \cellcolor{black!15}
    $0$
\\[0.8ex]
\hline
\end{tabular}
\caption{Canonicalized Poisson brackets among all constraints in
the Starobinsky model. Each entry gives~$\tt \{ row , column \} \approx entry$.
The shaded block denotes the second-class sector. All remaining
constraints are first-class.}
\label{FinalBrackets1}
\end{table}
It shows Hamiltonian and momentum constraints that are
first-class, whereas the traceless constraints~$\Phi^{ij}$
and~$\Psi^{ij}$ constitute a second-class pair. Therefore, in total we have~$N_{\rm can} \!=\! 24$ canonical 
variables,~$N_{\rm 1st} \!=\!4$ first-class constraints, 
and~$N_{\rm 2nd} \!=\! 10$ second-class constraints. This implies
that the count of the propagating degrees of 
freedom\footnote{The Dirac--Bergmann analysis distinguishes propagating degrees 
of freedom from constrained or gauge degrees of freedom by their role in the 
Cauchy problem. This does not mean that the constrained sector is physically 
empty: constrained variables may encode charges, boundary data, or other 
non-propagating observables; see the discussion in~\cite{Golovnev:2022rui}.}
is
\begin{equation}
N_{\rm phy} = \frac{1}{2} \Big(
    N_{\rm can} - 2 \!\times\! N_{\rm 1st} - N_{\rm 2nd}
    \Big)
    =
    3 \, ,
\end{equation}
corresponding to the standard result for the number of degrees of
freedom for~$f(R)$ theories. However, this counting is not valid
everywhere.

The crucial observation is that the character of the traceless
constraints changes when~$\rho \!\approx\! 0$, as is evident from
the second-class block in Table~\ref{FinalBrackets1}. At such
points the second-class pair degenerates, signaling the presence
of a singular surface in phase space. As in pure~$R^2$ gravity,
this degeneracy implies a discontinuous change in the perturbative
counting of degrees of freedom. In the present case, however, the
singular surface is not located at vanishing Ricci scalar. Instead,
it is determined by
\begin{equation}
\frac{\rho}{\sqrt{h}} \approx - 6 \Big( \frac{1}{\kappa^2} + 2 \alpha R \Big) \approx 0 \, .
\end{equation}
Thus, the Einstein-Hilbert term does not remove the singular
surface; it only shifts its location to
\begin{equation}
R \approx - \frac{1}{2\alpha\kappa^2} \, .
\label{CriticalSurface}
\end{equation}
In this sense, the singular behaviour found in pure~$R^2$ gravity
persists continuously in the Starobinsky model. What changes is
not the existence of the singular surface, but its position in
field space.

This result suggests that, in general~$f(R)$ theories, the
relevant singular surfaces should be associated with the
condition~$f'(R)\!=\!0$. We will establish this explicitly in
the remainder of this section.

\subsection{Convex and concave $f(R)$ models}
\label{subsec: Convex  and concave f(R) models}

We now extend the analysis from the Starobinsky model to the class
of $f(R)$ theories~(\ref{IntroAction}) for which~$f''(R)\!\neq\!0$
everywhere, i.e.~to theories without inflection points. This class
includes both convex and concave functions, and provides the
natural next step in generality between the special quadratic model
considered in the previous subsection and the completely
general~$f(R)$ theories treated below. For the Hamiltonian analysis,
it is convenient to rewrite the theory in an equivalent first-order
form by introducing an auxiliary scalar field~$\chi$,
\begin{equation}
S \bigl[ g_{\mu\nu}, \chi \bigr]
	=
	\int\! d^4x \, \sqrt{-g} \, 
	\Bigl[
	f(\chi) - f'(\chi) ( \chi - R )
	\Bigr]
    \, .
\label{ConcaveAuxiliaryRepresetation}
\end{equation}

As long as~$f''(\chi)\!\neq\!0$, this representation is
classically equivalent to the original one. Indeed, varying the
action with respect to~$\chi$ gives
\begin{equation}
\frac{\delta S}{\delta \chi}
    =
    - f''(\chi) (\chi - R)
    \approx 0
    \, 
\end{equation}
and therefore enforces~$\chi\!\approx\!R$. The assumption
that~$f''(R)\!\neq\!0$ is thus essential for the equivalence of the
two formulations. This is precisely why theories with inflection
points must be treated separately in the next subsection: when
~$f''(\chi)$ vanishes, the relation~$\chi\!\approx\!R$ is no
longer enforced in the same way, and the canonical structure may
change qualitatively.

Within the present class of theories, however, the auxiliary-field
representation is fully adequate and substantially simplifies the
intermediate steps of the constraint analysis.

\subsubsection{Canonical action}

After the ADM decomposition introduced in
Sec.~\ref{sec: ADM decomposition}, the
action~(\ref{ConcaveAuxiliaryRepresetation}) takes the form
\begin{equation}
S \bigl[ N, N_i, h_{ij}, \chi \bigr]
	= 
	\int\! d^4x \, N \sqrt{h} \, 
	\Bigl[
	f(\chi) - f'(\chi) \chi + f'(\chi) \Bigl( 2F + K^2 - K^{ij} K_{ij} + \mathcal{R} \Bigr)
	\Bigr]
	\, ,
\label{ADMaction}
\end{equation}
which provides a convenient starting point for the canonical
formulation and the subsequent constraint analysis.

To pass to canonical form, we proceed as in the Starobinsky case
and introduce an extended action by promoting the time-derivative
variables to independent velocity fields,
\begin{equation}
\dot{\chi} \longrightarrow Nv \, ,
\qquad \quad
K_{ij} \longrightarrow \mathcal{K}_{ij} \, ,
\qquad \quad
F_{ij} \longrightarrow \mathcal{F}_{ij} \, ,
\end{equation}
and by introducing the Lagrange
multipliers~$\pi^{ij}$,~$\rho^{ij}$, and~$\sigma$ to enforce
on-shell equivalence with the original formulation,
\begin{align}
\MoveEqLeft[2]
\mathcal{S}\bigl[ N, N_i, h_{ij}, \chi, v, \sigma, \mathcal{K}_{ij}, \pi^{ij}, \mathcal{F}_{ij}, \rho^{ij} \bigr]
    =
    \int\! d^4x \, \biggl\{
	N \sqrt{h} \, \Bigl[
	f(\chi) - f'(\chi) \chi 
\nonumber \\
&
    + f'(\chi) \Bigl( 2\mathcal{F} + \mathcal{K}^2 - \mathcal{K}^{ij} \mathcal{K}_{ij} + \mathcal{R} \Bigr)
	\Bigr]
    +
    \sigma (\dot{\chi} - Nv)
	+ \pi^{ij} \Bigl( \dot{h}_{ij} - 2 \nabla_{(i} N_{j)} 
	+ 2 N \mathcal{K}_{ij} \Bigr)
\nonumber \\
&
	+ \rho^{ij} \Bigl( \dot{\mathcal{K}}_{ij}
		+ N \mathcal{K}_{ik} {\mathcal{K}^k}_j
		- N^k \nabla_k \mathcal{K}_{ij}
	- 2 \mathcal{K}_{k(i} \nabla_{j)} N^k
	+ \nabla_i \nabla_j N + N \mathcal{F}_{ij} \Bigr)
    \biggr\}
    \, .
\label{ExtAct}
\end{align}

At this stage an important simplification occurs. Unlike in the
Starobinsky model, the field~$\mathcal{F}_{ij}$ appears only
linearly in the extended action~(\ref{ExtAct}). It therefore does
not need to be solved for algebraically; instead, it already plays
the role of a Lagrange multiplier. It is convenient to separate its
trace and traceless parts,
\begin{equation}
\mathcal{F}_{ij} \longrightarrow - \lambda_{ij} - \omega h_{ij} \, ,
\qquad
\text{such that} \qquad
\omega = - \mathcal{F}/3 \, ,
\qquad
\lambda_{ij} = - H_{ij}{}^{k\ell} \mathcal{F}_{k\ell} \, ,
\end{equation}
after which the action can be rewritten directly in canonical form
as
\begin{align}
\MoveEqLeft[3]
\mathscr{S}\bigl[ N, N_i, v, \omega, \lambda_{ij},
    \chi, \sigma, h_{ij}, \pi^{ij}, \mathcal{K}_{ij}, \rho^{ij} \bigr]
\nonumber \\
&
    =
    \int\! d^4x \, \Big[
    \sigma \dot{\chi}
    + \pi^{ij} \dot{h}_{ij}
    + \rho^{ij} \dot{\mathcal{K}}_{ij}
    -
	N \Big( \mathcal{H} + v \sigma
        + \omega \Xi + \lambda_{ij} \Phi^{ij} \Big)
    -
    N_i \mathcal{H}^i
    \Big].
\label{ConcaveCanonicalAction}
\end{align}
By suitable shifts of the multipliers, all explicit occurrences
of~$\rho^{ij}$ can be removed from the Hamiltonian and momentum
constraints. The resulting tentative Hamiltonian and momentum
constraints are
\begin{align}
\mathcal{H} ={}&
    \sqrt{h}
    \bigg[
    - 2 \mathcal{K}_{ij} \frac{\pi^{ij}}{\sqrt{h}}
        -
        f(\chi)
        +
        \Bigl(
        2 \nabla^2
        + \chi
        + 3\mathcal{K}^{ij} \mathcal{K}_{ij}
        - \mathcal{K}^2
        - \mathcal{R} 
        \Bigr)
        f'(\chi) 
    \bigg]
        \, ,
\label{ConcaveHamiltonianConstraint}
\\
\mathcal{H}^i ={}&
    \sqrt{h} 
    \bigg[ - 2 \nabla_{j} \Big( \frac{\pi^{ij}}{\sqrt{h}} \Big)
        +
        4 \nabla_j \big( f'(\chi)\mathcal{K}^{ij} \big)
        -
        2 f'(\chi) \nabla^i \mathcal{K}
        \bigg]
    \, .
\label{ConcaveMomentumConstraint}
\end{align}
In addition, the canonical action contains two scalar primary
constraints,
\begin{equation}
\sigma
\qquad\qquad\text{and}\qquad\qquad
\Xi = \rho + 6 \sqrt{h} f'(\chi) \, ,
\label{ConcaveScalarConstraint}
\end{equation}
as well as one traceless primary constraint,
\begin{equation}
\Phi^{ij} = H^{ij}{}_{k\ell} \rho^{k\ell} \, .
\label{ConcaveTransverseConstraint}
\end{equation}
This canonical formulation provides the starting point for the
Dirac--Bergmann analysis in the next subsection.

\subsubsection{Constraint analysis}

We now apply the Dirac--Bergmann algorithm to the canonical
formulation obtained in the previous subsection. The dynamics
following from the canonical action~(\ref{ConcaveCanonicalAction})
is generated by the total Hamiltonian, which in condensed notation
is
\begin{equation}
H_{\rm tot} = 
	\mathcal{H} \llbracket N \rrbracket
    +
    \sigma \llbracket Nv \rrbracket
    +
    \Xi \llbracket N\omega \rrbracket
    +
    \Phi^{ij} \llbracket N\lambda_{ij} \rrbracket
    +
    \mathcal{H}^i \llbracket N_i \rrbracket
    \, .
\end{equation}
The Poisson brackets among the primary constraints are summarized
in Table~\ref{ConcavePrimaryTable}. To present this algebra in a
compact form, we introduce three recurrent scalar combinations,
\begin{equation}
\Upsilon = \sqrt{h} \, f'(\chi)
\, ,
\qquad\quad
\Theta = \sqrt{h} f''(\chi)
\, ,
\qquad\quad
\Pi = \pi - \sqrt{h} \, f'(\chi) \mathcal{K}
\, ,
\end{equation}
and one recurrent traceless combination,
\begin{equation}
\Psi^{ij} = H^{ij}{}_{kl} \Big( \pi^{kl}
        -
        \sqrt{h} \, f'(\chi) \mathcal{K}^{kl} \Big)
    \, .
\label{PsiTraceless}
\end{equation}
%
\begin{sidewaystable}
\renewcommand{\arraystretch}{1.9}
\setlength\tabcolsep{0pt}
\centering
\begin{tabular}{ | w{c}{1.5cm} || w{c}{6.8cm} w{c}{6.2cm} w{c}{5.5cm} w{c}{2.cm} w{c}{2.5cm} |} 
\hline
	&
	$\mathcal{H} \llbracket s \rrbracket$
	&
	$\mathcal{H}^k \llbracket s_k \rrbracket$
    &
    $\sigma \llbracket s \rrbracket$
    &
    $\Xi \llbracket s \rrbracket$
	&
	$\Phi^{k\ell} \llbracket s_{k\ell} \rrbracket $
\\[0.8ex]
\hline\hline
	$\mathcal{H} \llbracket q \rrbracket $
	&
	$\begin{gathered}
	2 \Upsilon \Big\llbracket \mathcal{K}^{ij} 
        \nabla_i (s \nabla_j q - q \nabla_j s ) \Big\rrbracket
    \\
    +
    2 \Upsilon \Big\llbracket \nabla^i \mathcal{K} 
        (s \nabla_i q - q\nabla_i s) \Big\rrbracket
    \end{gathered}
    $
	&
	$
    \begin{gathered}
    \\[-1ex]
    2 \Psi^{ij} \Big\llbracket
    \nabla^k (q\mathcal{K}_{ij} s_k) 
    \!-\! \nabla_{j} (q\mathcal{K} s_i)
\\
    + 2q \mathcal{K}_{i}{}^k \nabla_j s_k
    \Big\rrbracket
    +
    \frac{4}{3} \Pi \big\llbracket \mathcal{K}^{ij} q\nabla_i s_j \big\rrbracket
\nonumber \\
    +
    2 \Upsilon \Big\llbracket
    3qs_k  \mathcal{K}_{ij} \nabla^k \mathcal{K}^{ij}
    +
    \mathcal{K}^{ij} \mathcal{K}_{ij} \nabla^k ( q s_k)
    \Big\rrbracket
\\
    +
    2 \Upsilon \Big\llbracket
    \mathcal{K}^{ij} \nabla_i ( \mathcal{K} q s_j)
    \!+\!
    \mathcal{R}^{ij} \nabla_i ( q s_{j} ) \Big\rrbracket
\nonumber \\
    +
    2 \Upsilon \Big\llbracket
    \nabla^j ( q \nabla^i \nabla_i s_j )
    -
    2 \nabla^i \nabla^j ( q \nabla_i s_j )
    \Big\rrbracket
\end{gathered}
    $
	&
	$\begin{gathered}
        \qquad
        \Theta \Big\llbracket s \big( 2\nabla^2 \!+\! \chi \!-\! \mathcal{K}^2 
        \qquad
        \\
        \qquad
        + 3\mathcal{K}^{ij} \mathcal{K}_{ij}
        \!-\! \mathcal{R} \big) q \Big\rrbracket
        \end{gathered}$
	&
	$-2\Pi \llbracket qs \rrbracket $
	&
	$-2 \Psi^{ij} \llbracket q s_{ij} \rrbracket $
\\[5ex]
	$\mathcal{H}^i \llbracket q_i \rrbracket $
	&
	$
    \begin{gathered}
    \\
    - 2 \Psi^{ij} \Big\llbracket
    \nabla^k (s\mathcal{K}_{ij} q_k) 
    \!-\! \nabla_{j} (s\mathcal{K} q_i)
\\
    + 2s \mathcal{K}_{i}{}^k \nabla_j q_k
    \Big\rrbracket
    -
    \frac{4}{3} \Pi \big\llbracket \mathcal{K}^{ij} s\nabla_i q_j \big\rrbracket
\\
    -
    2 \Upsilon \Big\llbracket
    3sq_k  \mathcal{K}_{ij} \nabla^k \mathcal{K}^{ij}
    +
    \mathcal{K}^{ij} \mathcal{K}_{ij} \nabla^k ( s q_k)
    \Big\rrbracket
\\
    -
    2 \Upsilon \Big\llbracket
    \mathcal{K}^{ij} \nabla_i ( \mathcal{K} s q_j)
    \!+\!
    \mathcal{R}^{ij} \nabla_i ( s q_{j} ) \Big\rrbracket
\\
    -
    2 \Upsilon \Big\llbracket
    \nabla^j ( s \nabla^i \nabla_i q_j )
    -
    2 \nabla^i \nabla^j ( s \nabla_i q_j )
    \Big\rrbracket
\end{gathered}
    $
	&
	$\begin{gathered}
    4 \Upsilon \Big\llbracket
        \nabla^k \Big( \mathcal{K}^{ij} 
            \big(
            q_k \nabla_i s_j
            \!-\! 
            s_k \nabla_i q_j \big) \Big) \Big\rrbracket
    \\
    +
    2 \Upsilon \Big\llbracket (\nabla^i \mathcal{K})
        \nabla^j \big( s_i q_j \!-\! q_i s_j \big) \Big\rrbracket
    \end{gathered}$
	&
	$\!\!\!\!
    - 2 \Theta \Big\llbracket 2 \mathcal{K}^{ij} s\nabla_i q_j
        \!+\! s q_i \nabla^i \mathcal{K} \Big\rrbracket $
	&
	$\ \ \ - 6 \Upsilon \big\llbracket \nabla^i (q_i s) \big\rrbracket $
	&
	$0$
\\[15ex]
	$\sigma \llbracket q \rrbracket $
	&
    $\begin{gathered}
        -
        \Theta \Big\llbracket q \big( 2\nabla^2 \!+\! \chi 
        \!-\! \mathcal{K}^2 
        \!+\! 3\mathcal{K}^{ij} \mathcal{K}_{ij}
        \!-\! \mathcal{R} \big) s \Big\rrbracket
        \end{gathered}$
    &
	$2 \Theta \Big\llbracket 2 \mathcal{K}^{ij} q\nabla_i s_j
        \!+\! q s_i \nabla^i \mathcal{K} \Big\rrbracket $
	&
	$0$
	&
	$-6\Theta \llbracket qs \rrbracket $
	&
	$0$
\\[5ex]
	$\Xi \llbracket q \rrbracket $
	&
	$2\Pi \llbracket qs \rrbracket $
	&
	$6 \Upsilon \big\llbracket \nabla^i (q s_i) \big\rrbracket $
	&
	$6\Theta \llbracket qs \rrbracket$
	&
	$0$
	&
	$0$
\\[4ex]
	$\Phi^{ij} \llbracket q_{ij} \rrbracket $
	&
	$2 \Psi^{ij} \llbracket q_{ij}s \rrbracket $
	&
	$0$
	&
	$0$
	&
	$0$
	&
	$0$
\\[3ex]
\hline
\end{tabular}
\caption{
Poisson brackets among the primary constraints
~(\ref{ConcaveHamiltonianConstraint})--(\ref{ConcaveTransverseConstraint})
in the auxiliary-field formulation of convex and concave~$f(R)$
theories. Each entry denotes
$\tt \{ row , column \} \approx entry$.
}
\label{ConcavePrimaryTable}
\end{sidewaystable}

A superficial inspection of Table~\ref{ConcavePrimaryTable} does
not immediately reveal the underlying constraint structure. In
particular, the tentative Hamiltonian and momentum
constraints,~$\mathcal{H}$ and~$\mathcal{H}^i$, do not have
vanishing on-shell brackets with the remaining primary constraints.
They therefore should not yet be interpreted as the final
first-class generators. This is not surprising: the canonical
action~(\ref{ConcaveCanonicalAction}) contains two scalar primary
constraints and one traceless primary constraint, and the
corresponding Lagrange multipliers can be shifted freely. Such
shifts redefine the combinations that ultimately play the role of
Hamiltonian and momentum constraints. At this stage, however, the
appropriate redefinitions are not yet apparent.

It is therefore more convenient to begin with the conservation of
the two scalar primary constraints in~(\ref{ConcaveScalarConstraint}),
\begin{align}
\dot{\sigma} \approx{}&
    \big\{ \sigma , H_{\rm tot} \big\}
    \approx
    \Theta
    \Big( \mathcal{K}^2 - 3 \mathcal{K}^{ij} \mathcal{K}_{ij}
        + \mathcal{R} - \chi - 2 \nabla^2 - 6\omega \Big) N
    +
    2 \Theta
    \Big(
    2 \mathcal{K}^{ij} \nabla_i N_j
    +
    N_i \nabla^i \mathcal{K}
    \Big)
    \, ,
\\
\dot{\Xi} \approx{}&
    \big\{ \Xi , H_{\rm tot} \big\}
    \approx
    2 \Pi N
    +
    6 \Theta \big( Nv - N_i \nabla^i \chi \big)
    \, .
\end{align}
Since~$f''(\chi)\!\neq\!0$ in the class of theories considered
here, these equations determine the two Lagrange multipliers
associated with the scalar constraints on shell:
\begin{align}
v \approx \overline{v}
    ={}&
    \frac{ N_i \nabla^i \chi }{N}
    - \frac{1}{3f''(\chi)} \Big( \frac{\pi}{\sqrt{h}}
        - f'(\chi) \mathcal{K} \Big)
    \, ,
\label{ConvexVSolution}
\\
\omega \approx \overline{\omega}
    ={}&
    \frac{1}{6} \Big( 
        \mathcal{K}^2 - 3 \mathcal{K}^{ij} \mathcal{K}_{ij}
        - \chi + \mathcal{R} \Big)
    -
    \frac{1}{3N} \Big( \nabla^2 N
        - 2 \mathcal{K}^{ij} \nabla_i N_j
        - N_i \nabla^i \mathcal{K}
        \Big)
    \, .
\label{ConvexLambdaSolution}
\end{align}

The conservation of the primary traceless constraint,
\begin{equation}
\dot{\Phi}^{ij} \approx \big\{ \Phi^{ij} , H_{\rm tot} \big\}
    \approx
    2 N \Psi^{ij}
\end{equation}
generates the secondary traceless constraint
\begin{equation}
\Psi^{ij} \approx 0 \, ,
\label{ConcavePhiSecondary}
\end{equation}
defined in~(\ref{PsiTraceless}). Once the two scalar multipliers
have been fixed on shell by~(\ref{ConvexLambdaSolution})
and~(\ref{ConvexVSolution}), and the secondary traceless
constraint~(\ref{ConcavePhiSecondary}) has been identified, the
consistency conditions for the remaining primary constraints
simplify substantially. In fact, the tentative Hamiltonian and
momentum constraints~(\ref{ConcaveHamiltonianConstraint})
and~(\ref{ConcaveMomentumConstraint}) are then conserved on shell,
\begin{equation}
\dot{\mathcal{H}} \approx 0 \, ,
\qquad
\dot{\mathcal{H}}{}^i \approx 0 \, . 
\end{equation}

Thus, requiring preservation of the primary constraints generates
only one secondary constraint. Its bracket with itself vanishes,
while its brackets with the primary constraints are
\begin{subequations}
\begin{align}
\big\{ \Psi^{ij} \llbracket q_{ij} \rrbracket , \mathcal{H} \llbracket s \rrbracket \big\}
    \approx{}&
    -
    \frac{2}{3} \Pi \big\llbracket H^{ijk\ell} q_{ij} 
        \mathcal{K}_{k\ell} s \big\rrbracket
    +
    2 \Upsilon\big\llbracket
        H^{ijk\ell} q_{ij}
        \mathcal{K}_{km}
        \mathcal{K}_\ell{}^m s \big\rrbracket
    -
    \Upsilon \big\llbracket H^{ijk\ell} q_{ij} 
        \mathcal{K}_{k\ell} \mathcal{K} s \big\rrbracket
\nonumber \\
&
    +
    \Upsilon \big\llbracket H^{ijk\ell} s
        ( \nabla_k\nabla_\ell - \mathcal{R}_{k\ell} ) q_{ij} \big\rrbracket
    +
    2 \Upsilon \big\llbracket H^{ijk\ell} 
        \nabla_k( q_{ij} \nabla_\ell s ) \big\rrbracket
    \, ,
\\
\big\{ \Psi^{ij} \llbracket q_{ij} \rrbracket ,
    \mathcal{H}^k \llbracket s_k \rrbracket \big\}
    \approx{}&
    -
    2 \Upsilon \big\llbracket 
    H^{ij}{}_{k\ell} q_{ij} s_m
        \nabla^m\mathcal{K}^{k\ell} \big\rrbracket
    -
    2 \Upsilon \big\llbracket H^{ij}{}_{k\ell} q_{ij}
        \mathcal{K}^{km} \nabla^\ell s_m \big\rrbracket
\nonumber \\
&
    -
    \Upsilon \big\llbracket
    H^{ij}{}_{k\ell} \mathcal{K}^{k\ell} \nabla^m( q_{ij} s_m )
    \big\rrbracket
    \, ,\\
\big\{ \Psi^{ij} \llbracket q_{ij} \rrbracket , \Xi \llbracket s \rrbracket \big\}
    \approx{}&
    0
    \, ,
\\
\big\{ \Psi^{ij} \llbracket q_{ij} \rrbracket , \sigma \llbracket s \rrbracket \big\}
    \approx{}&
    -
    \Theta \big\llbracket
    s q_{ij}
    H^{ij}{}_{k\ell} \mathcal{K}^{k\ell}
    \big\rrbracket
    \, ,
\\
\big\{ \Psi^{ij} \llbracket q_{ij} \rrbracket ,
    \Phi^{k\ell} \llbracket s_{k\ell} \rrbracket \big\}
    \approx{}&
    \Upsilon \big\llbracket H^{ijk\ell} q_{ij} s_{k\ell} \big\rrbracket
    \, .
\end{align}
\end{subequations}
These relations imply that conservation of the secondary traceless
constraint,
$\dot{\Psi}^{ij} \!\approx\! \big\{ \Psi^{ij} , H_{\rm tot} \big\} \!\approx\! 0$,
generates no further constraints. Instead, it fixes the traceless
Lagrange multiplier on shell,
\begin{align}
\MoveEqLeft[10]
\lambda_{ij} \approx \overline{\lambda}_{ij}
    =    
    H_{ij}{}^{k\ell}\bigg[
    \mathcal{R}_{k\ell}
    -
    \frac{\sqrt{h}}{\Upsilon}
        \nabla_k \nabla_\ell \frac{\Upsilon}{\sqrt{h}}
    -
    2 \mathcal{K}_{km}\mathcal{K}^{m}{}_{\ell}
    +
    \mathcal{K}_{k\ell}\mathcal{K}
    +
    \frac{\Pi}{3\Upsilon}\,\mathcal{K}_{k\ell}
\nonumber\\
&
    -
    \frac{\nabla_k \nabla_\ell N}{N}
    +
    \frac{N_m \nabla^m\mathcal{K}_{k\ell}}{N}
    +
    \frac{2 \mathcal{K}_{k}{}^{m}\nabla_\ell N_m}{N}
    \bigg]
    \, .
\end{align}

Having determined all generations of constraints and imposed their
conservation, we can now make the final constraint structure
manifest. This is achieved by shifting the scalar and traceless
Lagrange multipliers in~(\ref{ConcaveCanonicalAction}) by their
on-shell values,
\begin{equation}
v \longrightarrow v + \overline{v} \, ,
\qquad\qquad
\omega \longrightarrow \omega + \overline{\omega} \, ,
\qquad\qquad
\lambda_{ij} \longrightarrow \lambda_{ij} + \overline{\lambda}_{ij} \, .
\end{equation}
These shifts redefine the Hamiltonian and momentum constraints:
\begin{align}
\boldsymbol{\mathcal{H}} \equiv{}&
    \mathcal{H}
    -
    \frac{\Pi \sigma}{3\Theta}
    -
    \frac{\sqrt{h}}{6}
    \Bigl(
        2 \nabla^2
        + \chi
        + 3\mathcal{K}^{ij}\mathcal{K}_{ij}
        - \mathcal{K}^2
        - \mathcal{R}
    \Bigr) \frac{\Xi}{\sqrt{h}}
\nonumber \\
&\qquad
    -
    \sqrt{h}
    \biggl[
        \nabla_i \nabla_j
        - \mathcal{R}_{ij}
        + \frac{\sqrt{h}}{\Upsilon} \nabla_i \nabla_j 
            \Big( \frac{\Upsilon}{\sqrt{h}} \Big)
        + 2 \mathcal{K}_{ik}\mathcal{K}^{k}{}_{j}
        - \mathcal{K}_{ij}\mathcal{K}
        - \frac{\Pi}{3\Upsilon}\,\mathcal{K}_{ij}
    \biggr] \frac{\Phi^{ij}}{\sqrt{h}}
    \, ,
\\
\boldsymbol{\mathcal{H}}^i \equiv{}&
    \mathcal{H}^i
    +
    \sigma \nabla^i \chi
    +
    \frac{1}{3}\,\Xi \nabla^i \mathcal{K}
    -
    \frac{2}{3}\sqrt{h}\,
    \nabla_j\!\Bigl(
        \mathcal{K}^{ij}\frac{\Xi}{\sqrt{h}}
        +
        3 \mathcal{K}^{i}{}_{k}\frac{\Phi^{jk}}{\sqrt{h}}
    \Bigr)
    +
    \Phi^{jk}\nabla^i \mathcal{K}_{jk}
    \, .
\end{align}
We also shift the secondary traceless constraint,
\begin{equation}
\boldsymbol{\Psi}^{ij}
    =
    \Psi^{ij}
    +
    \frac{1}{6}H^{ij}{}_{k\ell}\mathcal{K}^{k\ell}\Xi
    \, ,
\end{equation}
so that the full constraint algebra, displayed in
Table~\ref{ConcaveFinalTable}, becomes block-diagonal and the
constraint structure is manifest.
\begin{table}[h!]
\renewcommand{\arraystretch}{2}
\setlength\tabcolsep{0pt}
\centering
\begin{tabular}{ | w{c}{1.5cm} || w{c}{2.cm} w{c}{2.cm} w{c}{2.3cm} w{c}{2.3cm} w{c}{2.9cm} w{c}{2.9cm} |} 
\hline
	&
	$\boldsymbol{\mathcal{H}} \llbracket s \rrbracket $
	&
	$\boldsymbol{\mathcal{H}}^k \llbracket s_k \rrbracket $
    &
    $\sigma \llbracket s \rrbracket $
    &
    $\Xi \llbracket s \rrbracket $
	&
	$\Phi^{k\ell} \llbracket s_{k\ell} \rrbracket $
	&
	$\boldsymbol{\Psi}^{k\ell} \llbracket s_{k\ell} \rrbracket $
\\
\hline\hline
	$\boldsymbol{\mathcal{H}} \llbracket q \rrbracket $
	&
	$0$
	&
	$0$
	&
	$0$
	&
	$0$
	&
	$0$
	&
	$0$
\\
	$\boldsymbol{\mathcal{H}}^i \llbracket q_i \rrbracket $
	&
	$0$
	&
	$0$
	&
	$0$
	&
	$0$
	&
	$0$
	&
	$0$
\\
	$\sigma \llbracket q \rrbracket $
	&
	$0$
	&
	$0$
	&
    \cellcolor{black!15}
	$0$
	&
    \cellcolor{black!15}
	$- 6 \Theta \llbracket qs \rrbracket $
	&
	$0$
	&
	$0$
\\
	$\Xi \llbracket q \rrbracket$
	&
	$0$
	&
	$0$
	&
    \cellcolor{black!15}
	$6 \Theta \llbracket qs \rrbracket$
	&
    \cellcolor{black!15}
	$0$
	&
	$0$
	&
	$0$
\\
	$\Phi^{ij} \llbracket q_{ij} \rrbracket $
	&
	$0$
	&
	$0$
	&
	$0$
	&
	$0$
	&
    \cellcolor{black!15}
	$0$
	&
    \cellcolor{black!15}
    \hskip-4mm
    $-\Upsilon \big\llbracket H^{ijkl}q_{ij}s_{k\ell} \big\rrbracket $
\\
	$\boldsymbol{\Psi}^{ij} \llbracket q_{ij} \rrbracket $
	&
	$0$
	&
	$0$
	&
	$0$
	&
	$0$
	&
    \cellcolor{black!15}
	$ \Upsilon\! \big\llbracket H^{ijkl}q_{ij}s_{k\ell} \big\rrbracket $
	&
    \cellcolor{black!15}
	$0$
\\[0.8ex]
\hline
\end{tabular}
\caption{
Canonicalized Poisson brackets among all constraints identified
in convex and concave~$f(R)$ theories. Each entry gives
$\tt \{ row , column \} \approx entry$. The shaded blocks denote
the second-class sectors. All remaining constraints are first-class.
}
\label{ConcaveFinalTable}
\end{table}
In this formulation, outside the singularities
of the constraint structure, we have~$N_{\rm can} \!=\!26$
canonical variables (not counting Lagrange 
multipliers),~$N_{\rm 1st} \!=\! 4$ first-class constraints, 
and~$N_{\rm 2nd} \!=\! 12$ second-class constraints, giving
the~$N\!=\!3$ propagating degrees of freedom.

The structure of Poisson brackets given in
Table~\ref{ConcaveFinalTable} clearly reveals that singular
behaviour in this model appears on the
surfaces~$\Upsilon \!=\! \sqrt{h} f'(\chi) \!\approx\!0$. On this
surface the bracket between the traceless pair of
constraints~$(\Phi^{ij},\Psi^{ij})$ vanishes, and the pair
therefore changes character from second-class to first-class. The
Hamilton equations descending from the canonical
action~(\ref{ConcaveCanonicalAction}) further imply
that~$\chi \!\approx \!R$, so that the singular surface can be
identified with~$f'(R)\!\approx\!0$. This also explains why there
is no singular surface associated with the scalar pair of
second-class constraints~$(\sigma,\Xi)$: by assumption of the
formulation, here we are considering only convex or concave models
for which~$\Theta \!\approx\! \sqrt{h} f''(R) \!\neq\!0$.

\subsection{General $f(R)$ models}
\label{subsec: General f(R) models}

We now turn to completely general~$f(R)$ theories, allowing in
particular for the presence of inflection points at which
~$f''(R)\!=\!0$. In this case, the auxiliary-field representation
used in the previous subsection,
Eq.~(\ref{ConcaveAuxiliaryRepresetation}), is no longer suitable,
since its equivalence to the original theory relies precisely on
the condition~$f''(R)\!\neq\!0$. Once this assumption is relaxed,
the relation between the auxiliary scalar and the Ricci scalar is
no longer enforced in the same way, and the canonical structure
must be reconsidered from the outset.

Nevertheless, a different auxiliary-field representation remains
available, namely
\begin{equation}
S \big[ g_{\mu\nu}, \chi, \zeta \big]
    =
    \int\! d^4x \, \sqrt{-g} \, \Big[
    f(\chi) + \zeta (\chi - R)
    \Big]
    \, .
\label{GeneralAuxiliary}
\end{equation}
Away from points where~$f''(\chi)$ vanishes, solving for~$\zeta$
on shell and substituting the result back into the action
reproduces the auxiliary-field formulation
in~(\ref{ConcaveAuxiliaryRepresetation}). In the present
subsection, however, our goal is precisely to analyze situations
in which~$f''(R)$ may vanish, so we work directly with the more
general representation~(\ref{GeneralAuxiliary}).

This formulation is sufficiently general to capture both types of
singular surfaces relevant for the full theory: those associated
with~$f'(R)\!=\!0$, already anticipated in the previous
subsections, and those associated with~$f''(R)\!=\!0$, which
cannot be treated within the framework of
Sec.~\ref{subsec: Convex  and concave f(R) models}. The purpose of
this subsection is therefore to determine how the Hamiltonian
constraint structure is modified in the fully general case, and to
identify the corresponding singular loci in phase space.

\subsubsection{Canonical formulation}

The ADM decomposition from Sec.~\ref{sec: ADM decomposition}
allows us to write the action~(\ref{GeneralAuxiliary}) as
\begin{equation}
S\big[ N, N_i, h_{ij}, \chi, \zeta \big]
    =
    \int\! d^4x \, N \sqrt{h} \, \Big[
    f(\chi)
    +
    \zeta \Big( \chi - 2F - K^2 + K^{ij} K_{ij} - \mathcal{R} \Big)
    \Big]
    \, .
\label{GeneralAuxiliaryADM}
\end{equation}
We pass to the canonical formulation by promoting the relevant
time-derivative variables to independent velocity fields,
\begin{equation}
\dot{\chi} \longrightarrow N v \, ,
\qquad \quad
K_{ij} \longrightarrow \mathcal{K}_{ij} \, ,
\qquad \quad
F_{ij} \longrightarrow \mathcal{F}_{ij} \, ,
\end{equation}
with~$K_{ij}$ and~$F_{ij}$ defined in~(\ref{Kdef})
and~(\ref{Fdef}). Introducing the corresponding Lagrange
multipliers~$\sigma, \pi^{ij}, \rho^{ij}$ to enforce on-shell
equivalence, we obtain the extended action
\begin{align}
\MoveEqLeft[4]
\mathcal{S}\big[ N, N_i, h_{ij}, \chi, \zeta, v, \sigma, \mathcal{K}_{ij}, \pi^{ij}, \mathcal{F}_{ij}, \rho^{ij} \big]
    =
    \int\! d^4x \, 
    \bigg\{
    N\sqrt{h} \, \Big[
    f(\chi)
    + \zeta \Big( \chi - 2\mathcal{F} - \mathcal{K}^2
        + \mathcal{K}^{ij} \mathcal{K}_{ij} - \mathcal{R} \Big)
    \Big]
\nonumber \\
&
    +
    \sigma (\dot{\chi} - Nv)
	+ \pi^{ij} \Bigl( \dot{h}_{ij} - 2 \nabla_{(i} N_{j)} 
	+ 2 N \mathcal{K}_{ij} \Bigr)
	+ \rho^{ij} \Bigl( \dot{\mathcal{K}}_{ij}
	+ N \mathcal{K}_{ik} {\mathcal{K}^k}_j
	- N^k \nabla_k \mathcal{K}_{ij}
\nonumber \\
&
	- 2 \mathcal{K}_{k(i} \nabla_{j)} N^k
	+ \nabla_i \nabla_j N + N \mathcal{F}_{ij} \Bigr)
    \bigg\}
    \, .
\end{align}

A useful feature of this formulation is that both~$\zeta$ and the
trace of~$\mathcal{F}_{ij}$ can be eliminated algebraically. The
relevant equations are
\begin{equation}
\frac{\delta \mathcal{S}}{\delta \zeta}
    =
    N\sqrt{h} \Big( \chi - 2\mathcal{F} - \mathcal{K}^2
        + \mathcal{K}^{ij} \mathcal{K}_{ij} - \mathcal{R} \Big)
        \approx 0 \, ,
\qquad
\frac{\delta \mathcal{S}}{\delta \mathcal{F}_{ij}}
    =
    N \rho^{ij} - 2N\sqrt{h} \, \zeta h^{ij}
    \approx 0
    \, ,
\end{equation}
which are solved by
\begin{equation}
\zeta \approx \overline{\zeta} = \frac{1}{6} \frac{\rho}{ \sqrt{h} } \, ,
\qquad\quad
\mathcal{F} \approx
    \overline{\mathcal{F}} =
    \frac{1}{2} \Big( \chi - \mathcal{K}^2
        + \mathcal{K}^{ij} \mathcal{K}_{ij} - \mathcal{R} \Big)
    \, .
\end{equation}
Substituting these solutions back into the extended action gives
the canonical action
\begin{align}
\MoveEqLeft[4]
\mathscr{S}\big[ N, N_i, v, \lambda_{ij}, \chi, \sigma, h_{ij}, \pi^{ij}, \mathcal{K}_{ij}, \rho^{ij} \big]
\equiv
\mathcal{S}\big[ N, N_i, h_{ij}, \chi, \zeta \!\to\! \overline{\zeta}, v, \sigma, \mathcal{K}_{ij}, \pi^{ij}, 
\mathcal{F}_{ij} \!\to\! \tfrac{1}{3} h_{ij} \overline{\mathcal{F}} \!-\! \lambda_{ij}, \rho^{ij} \big]
\nonumber \\
={}&
    \int\! d^4x \, 
    \Big[
    \sigma \dot{\chi}
	+ \pi^{ij} \dot{h}_{ij}
	+ \rho^{ij} \dot{\mathcal{K}}_{ij}
    - N \big( \mathcal{H} + v \sigma + \lambda_{ij} \Phi^{ij} \big)
    - N_i \mathcal{H}^i
    \Big]
    \, .
\label{GeneralCanonicalAction}
\end{align}
The corresponding tentative Hamiltonian and momentum constraints are
\begin{align}
\mathcal{H} ={}&
    \sqrt{h} \bigg[
    -
    2 \mathcal{K}_{ij} \frac{\pi^{ij} }{\sqrt{h}}
    -
    f(\chi)
    -
    \frac{1}{6}
        \Big( 2\nabla^2
            + \chi 
            + 3\mathcal{K}_{ij} \mathcal{K}^{ij}
            - \mathcal{K}^2
            - \mathcal{R} \Big)
            \frac{\rho}{\sqrt{h}}
    \bigg]
    \, ,
\\
\mathcal{H}^i ={}&
    \sqrt{h}
    \bigg[
    -
    2 \nabla_{j} \Big( \frac{\pi^{ij}}{\sqrt{h}} \Big)
    +
    \frac{ \nabla^i \mathcal{K} }{3} \frac{\rho}{\sqrt{h}}
    -
    \frac{2}{3} \nabla_j \Big( \mathcal{K}^{ij} \frac{\rho}{\sqrt{h}} \Big)
    \bigg]
    \, ,
\end{align}
while the remaining primary constraints are one scalar constraint
and one traceless constraint,
\begin{equation}
\sigma \, ,
\qquad\qquad
\Phi^{ij} = H^{ij}{}_{k\ell} \rho^{k\ell} \, .
\end{equation}
This completes the canonical formulation of the fully general
theory and provides the starting point for the Dirac--Bergmann
analysis in the next subsection.

\subsubsection{Constraint analysis}

We now apply the Dirac--Bergmann algorithm to the canonical action
of the fully general theory. To streamline the notation, we
introduce the following combinations,
\begin{equation}
\Upsilon = \sqrt{h} f'(\chi)+\frac{\rho}{6} \, ,
\qquad
\Theta = \sqrt{h} f''(\chi) \, ,
\qquad
\Pi = \pi + \frac{\rho \mathcal{K}}{6} \, ,
\qquad
\Psi^{ij} = H^{ij}{}_{k\ell}
    \Big( \pi^{k\ell} + \frac{\rho}{6} \mathcal{K}^{k\ell} \Big)
\, ,
\end{equation}
which allow the brackets among the primary constraints to be
written compactly in Table~\ref{PrimaryBracketsTable3}.
\begin{table}[h!]
\renewcommand{\arraystretch}{2}
\setlength\tabcolsep{0pt}
\centering
\begin{tabular}{ | w{c}{1.4cm} || w{c}{5.1cm} w{c}{4.8cm} w{c}{2.2cm} w{c}{2.4cm} |} 
\hline
	&
	$\mathcal{H} \llbracket s \rrbracket$
	&
	$\mathcal{H}^k \llbracket s_k \rrbracket$
    &
    $\sigma \llbracket s \rrbracket$
	&
	$\Phi^{k\ell} \llbracket s_{k\ell} \rrbracket$
\\
\hline\hline
	$\mathcal{H} \llbracket q \rrbracket$
	&
	$2\Psi^{ij} \big\llbracket s\nabla\!_i \nabla\!_j q 
        \!-\! q\nabla\!_i\nabla\!_j s \big\rrbracket$
	&
	$2 \Psi^{ij} \bigl\llbracket 2q \mathcal{K}_{i}{}^{k}
        \nabla_j s_k \!+\! q s_k \nabla^k \mathcal{K}_{ij} \bigr\rrbracket$
	&
    $-\Upsilon \llbracket qs \rrbracket$
    &
	$-2 \Psi^{ij} \llbracket s_{ij}q \rrbracket$
\\
	$\mathcal{H}^i \llbracket q_i \rrbracket$
	&
	$\ - 2 \Psi^{ij} \bigl\llbracket 2s \mathcal{K}_{i}{}^{k}
        \nabla_j q_k \!+\! s q_k \nabla^k \mathcal{K}_{ij} \bigr\rrbracket$
	&
	$0$
	&
    $0$
    &
	$0$
\\
	$\sigma \llbracket q \rrbracket$
	&
	$\Upsilon \llbracket qs \rrbracket $
	&
	$0$
	&
    $0$
    &
	$0$
\\
	$\Phi^{ij} \llbracket q_{ij} \rrbracket$
	&
	$2 \Psi^{ij} \llbracket q_{ij}s \rrbracket$
	&
	$0$
	&
    $0$
    &
	$0$

\\[0.8ex]
\hline
\end{tabular}
\caption{
Poisson brackets among the primary constraints of the fully
general~$f(R)$ theory. Each entry gives
$\tt \{ row , column \} \approx entry$.
}
\label{PrimaryBracketsTable3}
\end{table}

Table~\ref{PrimaryBracketsTable3} shows that preservation of the
primary constraints generates one scalar and one traceless
secondary constraint,
\begin{equation}
\Upsilon \approx 0 \, ,
\qquad\quad
\Psi^{ij} \approx 0
\, ,
\end{equation}
whose mutual brackets vanish weakly. The only nonvanishing brackets
between these secondary constraints and the primary constraints are
\begingroup
\allowdisplaybreaks
\begin{subequations}
\begin{align}
\big\{ \Upsilon \llbracket q \rrbracket , \mathcal{H}^i \llbracket s_i \rrbracket \big\}
    \approx{}&
    - \frac{1}{6} \rho \llbracket \nabla^i(qs_i) \rrbracket
    \, ,
\\
\big\{ \Upsilon \llbracket q \rrbracket , \mathcal{H} \llbracket s \rrbracket \big\}
    \approx{}&
    \frac{1}{3}\Pi \llbracket qs \rrbracket
    \, ,
\\
\big\{ \Upsilon \llbracket q \rrbracket , \sigma \llbracket s \rrbracket \big\}
    \approx{}&
    \Theta \llbracket qs \rrbracket \, ,
\\
\big\{ \Psi^{ij} \llbracket q_{ij} \rrbracket , \mathcal{H} \llbracket s \rrbracket \big\}
	\approx{}&
- \frac{1}{9} \rho \big\llbracket
        H^{ijkl} \big( 3 \mathcal{K}_{km} \mathcal{K}^m{}_l
            \!-\! \mathcal{K}_{kl} \mathcal{K} \big) q_{ij} s
        \big\rrbracket
    -
    \frac{1}{3} \pi \big\llbracket H^{ijkl} \mathcal{K}_{kl} q_{ij} s \big\rrbracket
\nonumber \\
&
    +
    \frac{1}{6} 
    \rho \big\llbracket
    H^{ijkl} s (\mathcal{R}_{kl} - \nabla_k \nabla_l ) q_{ij}
    \big\rrbracket
    -
    \frac{1}{3} \rho\big\llbracket H^{ijkl} \nabla_k \big( q_{ij} \nabla_l s \big) \big\rrbracket
    \ ,
\\
\big\{ \Psi^{ij} \llbracket q_{ij} \rrbracket , \mathcal{H}^k \llbracket s_k \rrbracket \big\}
	\approx{}&
    \frac{1}{3} \rho \big\llbracket H^{ijk\ell} q_{ij} \mathcal{K}_{k}{}^m \nabla_\ell s_m \big\rrbracket
    +
    \frac{1}{6} \rho \big\llbracket H^{ijk\ell} q_{ij} s_m \nabla^m \mathcal{K}_{k\ell} \big\rrbracket
    \, ,
\\
\big\{ \Psi^{ij}\llbracket q_{ij} \rrbracket , \Phi^{k\ell} \llbracket s_{k\ell} \rrbracket \big\} \approx{}&
    - \frac{1}{6} \rho \big\llbracket H^{ijk\ell} q_{ij} s_{k\ell} \big\rrbracket
    \, .
\end{align}
\end{subequations}
\endgroup

We next require conservation of the secondary constraints. This
gives
\begin{align}
\dot{\Upsilon} \approx{}&
    \frac{1}{3} N \Pi
    + \Theta N v
    +
    \frac{1}{6} \sqrt{h} \, N_i \nabla^i \Big( \frac{\rho}{\sqrt{h}} \Big)
    \approx 0
    \, ,
\\
\dot{\Psi}^{ij} \approx{}&
    \frac{ N H^{ijk\ell} }{6} \bigg[
    -
    2 \pi \mathcal{K}_{k\ell}
    +
    \rho \Big( \mathcal{R}_{k\ell} - 2  \mathcal{K}_{km} \mathcal{K}^m{}_\ell
            + \frac{2}{3} \mathcal{K}_{k\ell} \mathcal{K}
            - \frac{ \nabla_k \nabla_\ell N }{N} \Big)
    -
    \sqrt{h} \, \nabla_k \nabla_\ell \Big( \frac{\rho}{\sqrt{h}} \Big)
    \bigg]
\nonumber \\
&
    +
    \frac{ H^{ijk\ell} \rho}{6} \Big[
    2 \mathcal{K}_{k}{}^m \nabla_\ell N_m
    +
    N_m \nabla^m \mathcal{K}_{k\ell}
    \Big]
    -
    \frac{1}{6} \rho H^{ijk\ell} N \lambda_{k\ell}
    \approx 0 \, .
\end{align}
Rather than generating further constraints, these equations
determine the remaining Lagrange multipliers on shell:
\begin{align}
v \approx{}& 
    \overline{v}
    =
    -
    \frac{1}{6N\Theta}
    \bigg[
    2N\Pi
    +
    \sqrt{h} N_i \nabla^i \Big( \frac{\rho}{\sqrt{h}} \Big)
    \bigg]
    \, ,
\\
\lambda_{ij} \approx{}& \overline{\lambda}_{ij} =
    H_{ij}{}^{k\ell} \bigg[
    \mathcal{R}_{k\ell} - 2  \mathcal{K}_{km} \mathcal{K}^m{}_\ell
            + \frac{2}{3} \mathcal{K}_{k\ell} \mathcal{K}
            - \frac{ \nabla_k \nabla_\ell N }{N}
    -
    \frac{2 \pi \mathcal{K}_{k\ell}}{\rho}
\nonumber \\
&  \hspace{3cm}
    -
    \frac{\sqrt{h}}{\rho} \nabla_k \nabla_\ell \Big( \frac{\rho}{\sqrt{h}} \Big)
    +
    \frac{ 2 \mathcal{K}_{k}{}^m \nabla_\ell N_m
    +
    N_m \nabla^m \mathcal{K}_{k\ell} }{N} 
    \bigg]
    \, ,
\end{align}
which exhausts all generations of constraints.

The final step is to make the first-class structure manifest. This
is achieved by shifting the two Lagrange multipliers by their
on-shell values,
\begin{equation}
v \longrightarrow v + \overline{v} \, ,
\qquad\quad
\lambda_{ij} \longrightarrow \lambda_{ij} + \overline{\lambda}_{ij} \, ,
\end{equation}
which induces the following redefinition of the Hamiltonian and
momentum constraints:
\begin{align}
&
\boldsymbol{\mathcal{H}} = 
    \mathcal{H}
    +
    \bigg[
    \mathcal{R}_{ij} - 2  \mathcal{K}_{ik} \mathcal{K}^k{}_j
            + \frac{2}{3} \mathcal{K}_{ij} \mathcal{K}
            - \nabla_i \nabla_j
    -
    \frac{2 \pi \mathcal{K}_{ij} }{\rho}
    -
    \frac{\sqrt{h}}{\rho} \nabla_i \nabla_j \Big( \frac{\rho}{\sqrt{h}} \Big)
    \bigg]
    \Phi^{ij}
    -
    \frac{\Pi \sigma }{3 \Theta}
    \, ,
\\
&
\boldsymbol{\mathcal{H}}^i =
    \mathcal{H}^i
    +
    \Phi^{jk} \nabla^i \mathcal{K}_{jk}
    -
    2 \sqrt{h} \nabla_j \Big( \mathcal{K}^i{}_{k} \frac{ \Phi^{jk} }{ \sqrt{h} } \Big) 
    -\frac{\sigma}{6\Theta} \sqrt{h} \nabla^i \Big( \frac{\rho}{\sqrt{h}} \Big)
    \, .
\end{align}
With these redefinitions, the first-class character of
$\boldsymbol{\mathcal{H}}$ and~$\boldsymbol{\mathcal{H}}^i$
becomes transparent. The resulting constraint algebra is
summarized in Table~\ref{GeneralFinalTable}.
\begin{table}[h!]
\renewcommand{\arraystretch}{2}
\setlength\tabcolsep{0pt}
\centering
\begin{tabular}{ | w{c}{1.5cm} || w{c}{2.cm} w{c}{2.cm} w{c}{2.3cm} w{c}{2.3cm} w{c}{2.9cm} w{c}{2.9cm} |} 
\hline
	&
	$\boldsymbol{\mathcal{H}} \llbracket s \rrbracket$
	&
	$\boldsymbol{\mathcal{H}}^k \llbracket s_k \rrbracket$
    &
    $\sigma \llbracket s \rrbracket$
    &
    $\Upsilon \llbracket s \rrbracket$
	&
	$\Phi^{k\ell} \llbracket s_{k\ell} \rrbracket$
	&
	$\Psi^{k\ell} \llbracket s_{k\ell} \rrbracket$
\\
\hline\hline
	$\boldsymbol{\mathcal{H}} \llbracket q \rrbracket$
	&
	$0$
	&
	$0$
	&
	$0$
	&
	$0$
	&
	$0$
	&
	$0$
\\
	$\boldsymbol{\mathcal{H}}^i \llbracket q_i \rrbracket$
	&
	$0$
	&
	$0$
	&
	$0$
	&
	$0$
	&
	$0$
	&
	$0$
\\
	$\sigma \llbracket q \rrbracket$
	&
	$0$
	&
	$0$
	&
    \cellcolor{black!15}
	$0$
	&
    \cellcolor{black!15}
	$-\Theta \llbracket qs \rrbracket$
	&
	$0$
	&
	$0$
\\
	$\Upsilon \llbracket q \rrbracket$
	&
	$0$
	&
	$0$
	&
    \cellcolor{black!15}
	$ \Theta \llbracket qs \rrbracket$
	&
    \cellcolor{black!15}
	$0$
	&
	$0$
	&
	$0$
\\
	$\Phi^{ij} \llbracket q_{ij} \rrbracket$
	&
	$0$
	&
	$0$
	&
	$0$
	&
	$0$
	&
    \cellcolor{black!15}
	$0$
	&
    \cellcolor{black!15}
    $\frac{1}{6} \rho \big\llbracket H^{ijk\ell} q_{ij} s_{k\ell} \big\rrbracket$
\\
	$\Psi^{ij} \llbracket q_{ij} \rrbracket$
	&
	$0$
	&
	$0$
	&
	$0$
	&
	$0$
	&
    \cellcolor{black!15}
	$ -\frac{1}{6} \rho \big\llbracket H^{ijk\ell} q_{ij} s_{k\ell} \big\rrbracket$
	&
    \cellcolor{black!15}
	$0$
\\[0.8ex]
\hline
\end{tabular}
\caption{
Canonicalized Poisson brackets among all constraints identified
in general~$f(R)$ theories. Each entry gives
$\tt \{ row , column \} \approx entry$. The shaded blocks denote
the second-class sectors. All remaining constraints are first-class.
}
\label{GeneralFinalTable}
\end{table}
Outside the possible singular surfaces, 
in this formulation we again have~$N_{\rm can} \!=\!26$
canonical variables (not counting Lagrange 
multipliers),~$N_{\rm 1st} \!=\! 4$ first-class constraints, 
and~$N_{\rm 2nd} \!=\! 12$ second-class constraints, giving
the~$N\!=\!3$ propagating degrees of freedom.

The structure displayed in Table~\ref{GeneralFinalTable} makes the
two singular surfaces of the general theory apparent. The first is
located at~$\rho\!\approx\!0$, where the traceless second-class
pair~$(\Phi^{ij},\Psi^{ij})$ changes character to first-class.
The second is located at~$\Theta \!=\! \sqrt{h} f''(\chi)\!\approx\!0$,
where the scalar pair of constraints~$(\sigma,\Upsilon)$ changes
character from second-class to first-class. Further insight into
these singular surfaces is obtained from the equations of motion
following from the canonical action~(\ref{GeneralCanonicalAction}),
which imply $\rho \!\approx\! - 6 \sqrt{h} f'(\chi)$ 
and $\chi \!\approx\! R $. Thus, the singular surfaces are located 
at~$f'(R)\!=\!0$ and~$f''(R)\!=\!0$, respectively.

\section{Perturbations around singular-surface solutions}
\label{sec: Perturbations around singular-surface solutions}

In this section we consider~$f(R)$ models that admit exact solutions 
satisfying~$f'(R)\!=\!0$ everywhere. The existence conditions for
such solutions were discussed
in~\cite{Casado-Turrion:2023rni,Casado-Turrion:2024esi}.
Moreover, the covariant equations of motion~(\ref{CovariantEOM})
imply that, on such solutions, one must also have
$f(R)\!=\!0$. We study linear perturbations around these
backgrounds, denoting background quantities by an overline and
perturbations by~$\delta$.

According to the Hamiltonian constraint analysis of the previous
section, these backgrounds lie on a singular surface in phase
space, where the constraint structure changes discontinuously.
One therefore expects the linearized constraint structure to
reflect this singular behaviour. This is precisely what happens
in pure~$R^2$ gravity, where perturbations around backgrounds
with~$R\!=\!0$ have an empty linearized spectrum despite the presence
of propagating degrees of freedom in the full nonlinear
theory~\cite{Hell:2023mph,Golovnev:2023zen,Karananas:2024hoh,Barker:2025gon}.
The emptiness of the linearized spectrum around maximally
symmetric spacetimes corresponding to~$f'(R)\!=\!0$
and~$f(R)\!=\!0$ has already been established
in~\cite{Casado-Turrion:2024esi}. Here we extend this result beyond
the maximally symmetric case, showing that the same mechanism
applies to arbitrary exact backgrounds of general~$f(R)$ models
admitting solutions with~$f(R)\!=\!0$ and~$f'(R)\!=\!0$, of which
the theory~$f(R)\!=\!R^2$ is a special case.

\medskip

We use the most general canonical formulation from
Sec.~\ref{subsec: General f(R) models}, since it is capable of
describing solutions with~$f''(R)\!=\!0$ and contains the more
special cases as limits. For solutions satisfying
~$f(\overline{R})\!=\!0$ and~$f'(\overline{R})\!=\!0$, the Hamilton
equations generated by the canonical 
action~(\ref{GeneralCanonicalAction})
imply the following for the background canonical variables:
\begin{equation}
\overline{\chi} = \overline{R} \, ,
\qquad
f(\overline{\chi}) = 0 \, ,
\qquad
\overline{\rho} = - 6 \sqrt{\overline{h}} \, 
    f'(\overline{\chi}) = 0 \, ,
\qquad
\overline{\pi}^{ij} = 0 \, ,
\qquad
\overline{\sigma} = 0 \, ,
\qquad
\overline{\rho}{}^{ij} = 0 \, .
\end{equation}
We perturb the canonical variables around such a background as
\begin{equation}
\chi \to \overline{\chi} + \delta\chi \, ,
\quad
\sigma \to \delta \sigma \, ,
\quad
h_{ij} \to \overline{h}_{ij} + \delta h_{ij} \, ,
\quad
\pi^{ij} \to \delta \pi^{ij} \, ,
\quad
\mathcal{K}_{ij} \to \overline{\mathcal{K}}_{ij}
    + \delta \mathcal{K}_{ij} \, ,
\quad
\rho^{ij} \to \delta \rho^{ij}
\, .
\end{equation}
The corresponding linearized primary constraints are
\begingroup
\allowdisplaybreaks
\begin{subequations}
\begin{align}
\mathcal{H}_{\scr (1)} ={}&
    \sqrt{\overline{h}}
    \bigg[
    -
    2 \overline{\mathcal{K}}_{ij}
        \frac{\delta\pi^{ij}}{\sqrt{\overline{h}}}
    -
    \frac{1}{6}
    \Big( 2 \overline{\nabla}{}^2
        + \overline{\chi}
        + 3 \overline{\mathcal{K}}^{ij} \overline{\mathcal{K}}_{ij}
        - \overline{\mathcal{K}}{}^2 - \overline{\mathcal{R}} \Big)
    \frac{\delta \rho}{\sqrt{\overline{h}}}
    \bigg]
    \, ,
\\
\mathcal{H}^i_{\scr (1)} ={}&
    \sqrt{\overline{h}}
    \bigg[
    -
    2 \overline{\nabla}_{j}
        \Big( \frac{\delta \pi^{ij}}{\sqrt{\overline{h}}} \Big)
    +
    \frac{ \overline{\nabla}{}^i \overline{\mathcal{K}} }{3}
        \frac{\delta \rho}{\sqrt{\overline{h}}}
    -
    \frac{2}{3} \overline{\nabla}_j \Big( \overline{\mathcal{K}}^{ij}
        \frac{\delta \rho}{\sqrt{\overline{h}}} \Big)
    \bigg]
    \, ,
\\
\sigma_{\scr(1)} ={}& 
    \delta \sigma \, ,
\qquad\qquad
\Phi^{ij}_{\scr (1)} =
    \overline{H}{}^{ij}{}_{k\ell} \delta \rho^{k\ell}
    \, .
\end{align}
\end{subequations}
\endgroup
Their conservation generates linearized secondary constraints
\begin{equation}
\Upsilon_{\scr (1)}
    = \sqrt{\overline{h}} \, f''(\overline{\chi}) \delta \chi + \frac{\delta \rho}{6}
    \, ,
\qquad\qquad
\Psi^{ij}_{\scr (1)} =
    \overline{H}{}^{ij}{}_{k\ell} \Big( \delta \pi^{k\ell}
    + \frac{\delta \rho}{6} \overline{\mathcal{K}}{}^{k\ell} \Big)
    \, ,
\end{equation}
the conservation of which generates no further constraints.

This set of linearized constraints requires careful interpretation. 
First, the Hamiltonian and momentum constraints depend only on the 
two scalar variables,~$\delta \pi \!=\! \overline{h}_{ij} \delta \pi^{ij}$ and~$\delta\rho \!=\! \overline{h}_{ij} \delta \rho^{ij}$,
after imposing the secondary constraints,
\begin{align}
\mathcal{H}_{\scr (1)} \approx{}&
    \sqrt{\overline{h}} \bigg[
    -
    \frac{2\overline{\mathcal{K}}}{3} \frac{\delta \pi }{\sqrt{\overline{h}}}
    -
    \frac{1}{6}
        \Big( 2\overline{\nabla}{}^2
            + \overline{\chi} 
            + \overline{H}{}^{ijk\ell} \overline{\mathcal{K}}_{ij} \overline{\mathcal{K}}_{k\ell}
            - \overline{\mathcal{R}} \Big)
            \frac{\delta \rho}{\sqrt{\overline{h}}}
    \bigg]
    \, ,
\label{HamPert}
\\
\mathcal{H}^i_{\scr (1)} \approx{}&
    \sqrt{\overline{h}}
    \bigg[
    -
    \frac{2}{3} \overline{\nabla}{}^i \Big( \frac{\delta \pi}{\sqrt{\overline{h}}} \Big)
    -
    \frac{ \overline{\mathcal{K} } }{9}
    \overline{\nabla}{}^i
        \Big( \frac{\delta \rho}{\sqrt{\overline{h}}} \Big)
    +
    \frac{2}{9} \frac{\delta \rho}{\sqrt{\overline{h}}}
        \overline{\nabla}{}^i \overline{\mathcal{K}}
    -
    \frac{1}{3} \overline{\nabla}{}_j \Big( \overline{\mathcal{K}}{}^{ij} 
        \frac{\delta \rho}{\sqrt{\overline{h}}} \Big)
    \bigg]
    \, .
\label{MomPert}
\end{align}
In particular, the transverse part of the momentum constraint,
\begin{equation}
\big( \overline{\nabla}_j \overline{\nabla}{}^i
    - \delta^i_j \overline{\nabla}{}^2 \big) 
    \frac{\mathcal{H}^j_{\scr (1)}}{\sqrt{\overline{h}}}
    =
    \frac{2}{3}
    \overline{\nabla}_j 
    \bigg[
    \overline{\nabla}{}^{[i}
        \Big( \frac{\delta \rho}{\sqrt{\overline{h}}} \Big)
        \overline{\nabla}{}^{j]} \overline{\mathcal{K} }
        \bigg]
    -
    \frac{1}{3} 
    \big( \overline{\nabla}_j \overline{\nabla}{}^i
        - 
        \delta^i_j \overline{\nabla}{}^2
        \big)
    \overline{\nabla}{}_k
    \Big( \overline{\mathcal{K}}{}^{kj} 
        \frac{\delta \rho}{\sqrt{\overline{h}}} \Big)
    \, ,
\label{longitudinal}
\end{equation}
depends only on~$\delta\rho$, so that it
degenerates to a single scalar condition, which may be represented
as~$\delta \rho \!\approx \!0$. Once this condition is imposed,
both the longitudinal part of the momentum constraint and the
Hamiltonian constraint reduce to a single scalar condition,
represented by~$\delta\pi \!\approx \!0$. Thus, the Hamiltonian
and momentum constraints in~(\ref{HamPert}) and~(\ref{MomPert})
account for only two first-class constraints, not four as would be
inferred naively. This conclusion persists even if the transverse
part of the momentum constraint~(\ref{longitudinal}) vanishes
identically on account of background fields.

Furthermore, the two linearized traceless constraints have
vanishing Poisson brackets, and therefore change character: they
become first-class at linear order, in contrast to their
second-class character in the full theory. Lastly, the character
of the two scalar constraints depends on whether the background is
also an inflection point of~$f(R)$:
\begin{itemize}
\item
In the first case we have~$f''(\overline{\chi}) \!\neq\! 0$,
which implies two second-class constraints~$\delta\sigma \!\approx\! 0$
and~$\delta \chi \!\approx\! 0$. The counting of linearized degrees of
freedom in this case is then:
\begin{equation}
N_{\rm phy} = \frac{1}{2} \Big( 26 - 2 \!\times\! 12 - 2 \Big) = 0 \, .
\end{equation}

\item
In the second case we have $f''(\overline{\chi}) \!=\! 0$. In this
case the secondary scalar constraint~$\Upsilon_{\scr (1)}$ reduces
to the condition already contained in the
Hamiltonian--momentum sector, while the primary
constraint~$\delta\sigma \!\approx \!0$ becomes first-class.
Therefore, the counting of the number of degrees of freedom
proceeds differently,
\begin{equation}
N_{\rm phy} = \frac{1}{2} \Big( 26 - 2 \!\times\! 13 - 0 \Big) = 0 \, .
\end{equation}
\end{itemize}

In both cases the result is the same: the linearized spectrum
around backgrounds satisfying~$f(\overline{R}) \!=\! 0$ and
$f'(\overline{R}) \!=\! 0$ is empty. This differs discontinuously from
the generic nonlinear theory, which propagates three degrees of
freedom. The disappearance of the linearized modes therefore does
not mean that the full theory loses its degrees of freedom on
nearby configurations; rather, it shows that the chosen background
lies on a singular surface of phase space where the perturbative
description degenerates.

This is directly analogous to the behaviour of perturbations around
the~$R\!=\!0$ surface in pure~$R^2$ gravity. It signals the strong
coupling of perturbative variables and raises the dynamical
question of whether such singular surfaces can be approached or
crossed by regular phase-space trajectories. We turn to this
question in the following section.

\section{Perturbations near singular-surface crossings}
\label{sec: Perturbations near singular-surface crossings}

In the previous section we considered exact backgrounds that lie
entirely on a singular surface of phase space. A different and more
dynamical situation arises when a regular background trajectory
approaches or crosses such a surface during its evolution. This
possibility was already encountered in pure~$R^2$ gravity, where
the cosmological phase space contains trajectories that cross the
singular surface~$R\!=\!0$~\cite{Barker:2025gon}. The perturbative
analysis performed there, however, concerned backgrounds lying on
the singular surface for all time, rather than perturbations around
trajectories that cross it dynamically.

The analysis of such crossings is substantially more delicate. One
must first identify backgrounds whose evolution reaches the
singular surface, and then determine how the perturbative
constraint structure behaves as the crossing is approached. These
two issues are not completely independent: the regularity of the
background evolution does not by itself guarantee that the
linearized perturbations evolve regularly through the same point.
For this reason, we restrict attention in this section to a
specific and tractable setting: FLRW backgrounds in the Starobinsky
model introduced in Sec.~\ref{subsec: Starobinsky model}. In the
following subsection we show that a large set of cosmological
phase-space trajectories crosses the surface~$f'(R)\!=\!0$. We then
analyze the constraint structure of perturbations around such
backgrounds in the vicinity of the crossing.

\subsection{Cosmological phase space analysis}
\label{sec: Cosmological phase space analysis}

Before turning to perturbations, we first show that the singular 
surface identified in the Hamiltonian analysis can be reached by 
regular cosmological evolution. We focus on the Starobinsky model, 
for which the singular surface is located 
at~$R\!=\!-1/(2\alpha\kappa^2)$.
The question is whether cosmological trajectories can pass through 
this value of the Ricci scalar. This provides the background setting 
for the perturbative analysis in the next subsection.

We restrict attention to spatially flat FLRW geometries,
\begin{equation} ds^2 = -dt^2 + a^2(t)d\vec{x}^{\,2} \, , 
\end{equation}
with Hubble rate~$H \!=\!\dot a/a$. For the Starobinsky action 
introduced in Sec.~\ref{subsec: Starobinsky model}, the only 
dimensionful parameter relevant for the homogeneous dynamics 
is~$\beta \!=\! \alpha\kappa^2$. The equations of motion can be 
written as
\begin{equation}
\dot{H} = \frac{R}{6} - 2H^2 
\, , 
\qquad \quad
H \dot{R} = \frac{R^2}{12} - R H^2 - \frac{H^2}{2 \beta }
\, , 
\qquad \quad
\ddot{R} = H\dot{R} - 2 \dot{H}\Bigl( R + \frac{1}{2\beta } \Bigr) 
\, . 
\label{StarCosmoEquations}
\end{equation}
The first equation is the definition of the Ricci scalar in a spatially 
flat FLRW spacetime, while the remaining equations are the independent 
cosmological equations of motion.

There is no unique or generally preferred choice of variables for 
dynamical system formulation of the cosmological background (e.g.~\cite{Carloni:2006mr,deSouza:2007zpn,Carloni:2007eu,Carloni:2007br,Odintsov:2017tbc,Chakraborty:2021mcf,Carloni:2004kp,Capozziello:1993xn,Alho:2026qxz,Carloni:2015jla}) so here we choose the variables best suited
for our purpose.
Since the theory contains a single dimensionful scale, it is natural to
introduce dimensionless variables by normalizing with~$\beta$. This is 
not possible in the pure~$R^2$ limit, where no scale is available, but 
for the present theory it provides a convenient parametrization of the 
phase space. We therefore define
\begin{equation}
X = |\beta|^{1/2} H \, ,
\qquad\quad
Y = |\beta|  R + \frac{b}{2} \, ,
\qquad\quad
Z = |\beta|^{3/2} \dot{R} \, ,
\end{equation}
where~$b \!=\! {\rm sgn}(\beta)$. The shift in the definition of~$Y$ 
is chosen so that the singular surface is located at~$Y\!=\!0$. We 
also introduce a dimensionless time variable~$T\!=\! t/|\beta|^{1/2}$. 
In terms of these variables, the system in~(\ref{StarCosmoEquations}) 
becomes
\begin{align}
\frac{dX}{dT}
    ={}&
    \frac{Y }{6}
    -
    2X^2
    -
    \frac{b}{12}
    \, ,
\label{eq1}
\\
\frac{dY}{dT}
    ={}&
    \frac{1}{12X}
    \Big( Y - \frac{b}{2} \Big)^{\!2}
    -
    X Y \, ,
\label{eq2}
\\
\frac{dZ}{dT}
    ={}&
    XZ
    -
    2 Y
    \Big(
    -
    2X^2
    +
    \frac{Y }{6}
    -
    \frac{b}{12} \Big)
    \, ,
\label{eq3}
\end{align}
which should be supplemented by
\begin{equation}
\frac{d Y}{dT} = Z \, .
\label{eq4}
\end{equation}
In this way, all choices of the two parameters~($\alpha,\beta$) are reduced to two 
discrete systems, distinguished only by the sign of~$\beta$. This is 
expected, since the overall normalization of the action does not 
affect the dynamics, while the relative sign between the two terms in 
the action does.

The dynamical system is in essence two-dimensional, but this two-dimensional 
surface is curved and embedded in a three-dimensional space. 
One may therefore either visualize the full three-dimensional embedding, 
or study a number of two-dimensional projections, as was done 
in~\cite{Barker:2025gon}. In the following, we focus on the latter, 
since they already suffice to show that the singular surface is 
generically crossed by cosmological trajectories.
This can also be inferred more formally from the system of 
equations above. In particular, specializing Eq.~(\ref{eq2}) to the singular
surface~$Y\!=\!0$ yields for the transverse 
derivative~$dY/dT \!\xrightarrow{Y \to 0}\! 1/(48X)$.
Thus, the surface~$Y=0$ does not act as an endpoint or barrier
for the homogeneous and isotropic evolution.

\begin{figure}[h!]
\includegraphics[width=15.8cm]{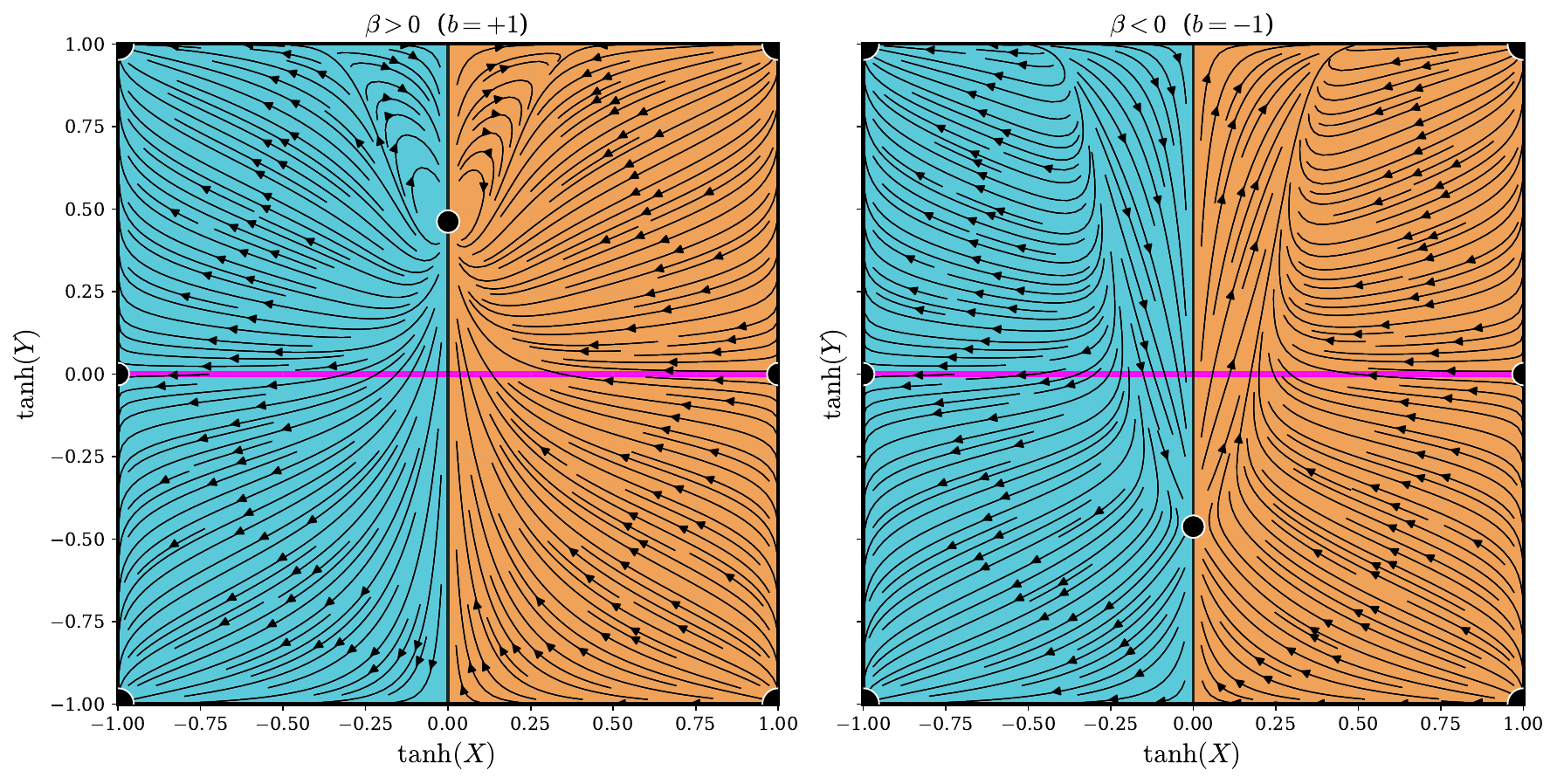}
\vskip-3mm
\caption{
Phase space flow corresponding to the dynamical system given 
in~(\ref{DynamicalSystem1}), with the left panel corresponding 
to~$\beta\!>\!0$, and the right panel corresponding to~$\beta\!<\!0$. 
The light blue and light orange colours signify disconnected sectors of 
the phase space, each with a definite sign of the Hubble rate.
Magenta line denotes the singular surface~$Y\!=\!0$, which is crossed 
without obstructions by a considerable portion of the phase space flow 
lines from both sectors. The black dots denote fixed points.}
\label{PhaseSpace1}
\end{figure}

\paragraph{First projection: $X$ and $Y$.}
A first projection is obtained by taking Eqs.~(\ref{eq1}) and~(\ref{eq2}) as
the defining system, while the third variable~$Z$ is determined by
Eq.~(\ref{eq4}),
\begin{equation}
\frac{dX}{dT}
    =
    \frac{Y }{6}
    -
    2X^2
    -
    \frac{b}{12}
    \, ,
\qquad\qquad
\frac{dY}{dT}
    =
    \frac{1}{12X}
    \Big( Y - \frac{b}{2} \Big)^{\!2}
    -
    X Y \, .
\label{DynamicalSystem1}
\end{equation}
The phase space flow corresponding to these equations is given in Fig.~\ref{PhaseSpace1}.
This projection already makes several qualitative features apparent. In
particular, the sign of~$X$, and hence the sign of the Hubble rate, is preserved
along the flow. At the same time, the singular surface~$Y=0$ does not in
general act as an impenetrable barrier: the stream lines indicate that it is
typically crossed during the evolution. The swirling structure visible in the
plot suggests a further subdivision of the flow into sectors, whose detailed
analysis is not pertinent to this work.
%

\begin{figure}[h!]
\includegraphics[width=16cm]{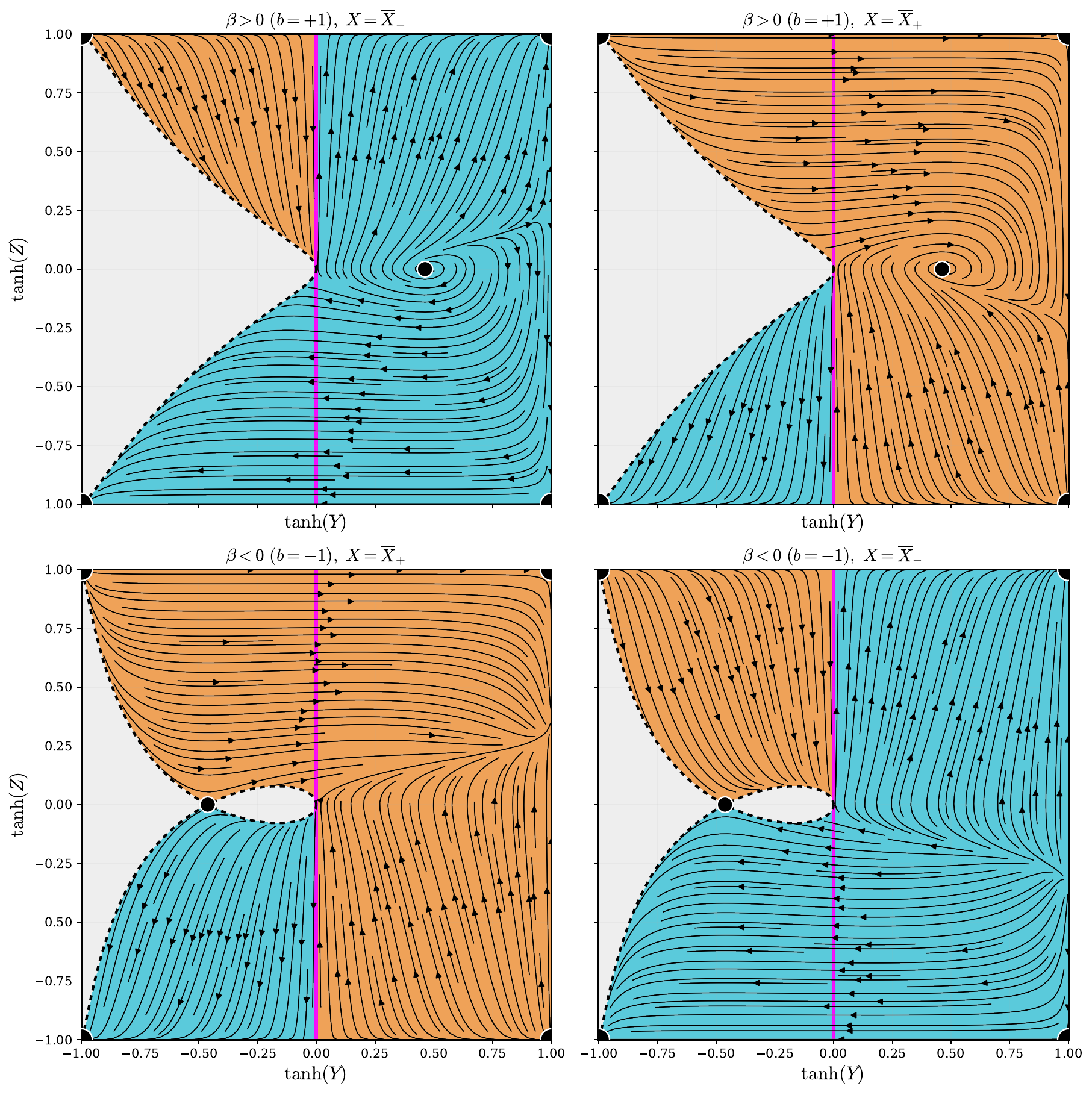}
\caption{
Phase space flow corresponding to the dynamical system given 
in~(\ref{DynamicalSystem2}), with the top panels corresponding 
to~$\beta\!>\!0$, and the bottom panels corresponding to~$\beta\!<\!0$. 
The light blue and light orange colours signify disconnected sectors of 
the phase space, each with a definite sign of the Hubble rate,
corresponding to the equally coloured sectors in Fig.~\ref{PhaseSpace1}.
The dashed black lines denote where the two top panels, and the two 
bottom panels are glued together, and the stream lines are seen to 
continue smoothly between panels. Magenta lines denote the singular 
surface~$Y\!=\!0$, which is crossed without obstructions by a 
considerable portion of the phase space flow lines from both sectors.
The black dots denote fixed points.}
\label{PhaseSpace2}
\end{figure}

\paragraph{Second projection: $Y$ and $Z$.}
A complementary picture is obtained by solving Eqs.~(\ref{eq2}) and~(\ref{eq4})
for~$X$ in terms of~$Y$ and~$Z$,
\begin{equation}
X = \overline{X}_{\pm}(Y,Z)
    =
    \frac{1}{2Y} \bigg[
    - Z \pm \sqrt{ Z^2 + \frac{ Y }{3} \Big( Y \!-\! \frac{b}{2} \Big)^{\!2} } \
    \bigg]
    \, .
\end{equation}
The resulting two-dimensional dynamical system is then
\begin{equation}
\frac{dY}{dT} = Z \, ,
\qquad\qquad
\frac{dZ}{dT} = 
    -
    3\overline{X}_{\pm}Z
    -
    \frac{b}{6} Y
    +
    \frac{1}{12}
    \, .
\label{DynamicalSystem2}
\end{equation}
The phase space flow corresponding to this system is given in Fig.~\ref{PhaseSpace2}.
This representation resolves part of the structure that is compressed in the
first projection. In particular, the crossing of~$Y=0$ is now seen from the perspective of the
Ricci scalar and its velocity, rather than from that of the Hubble expansion.
%

\paragraph{Third projection: $X$ and $Z$.}
Finally, one may instead eliminate~$Y$ by solving Eqs.~(\ref{eq2}) and
(\ref{eq4}) algebraically,
\begin{equation}
Y = \overline{Y}_{\pm}(X,Z) =
    6 \bigg[ X^2 + \frac{b}{12} 
        \pm \sqrt{ \frac{X}{6} \Big( 2Z + 6X^3 + b X \Big) } \ \bigg]
        \, .
\end{equation}
thus producing the following two-dimensional system
\begin{equation}
\frac{dX}{dT} = 
    \frac{\overline{Y}_{\pm}}{6}
    -
    2X^2
    -
    \frac{b}{12}
    \, ,
\qquad\quad
\frac{dZ}{dT} = 
    -
    3XZ
    -
    \frac{b}{6} \overline{Y}_{\pm}
    +
    \frac{1}{12}
    \, .
\label{DynamicalSystem3}
\end{equation}
\begin{figure}[h!]
\includegraphics[width=16cm]{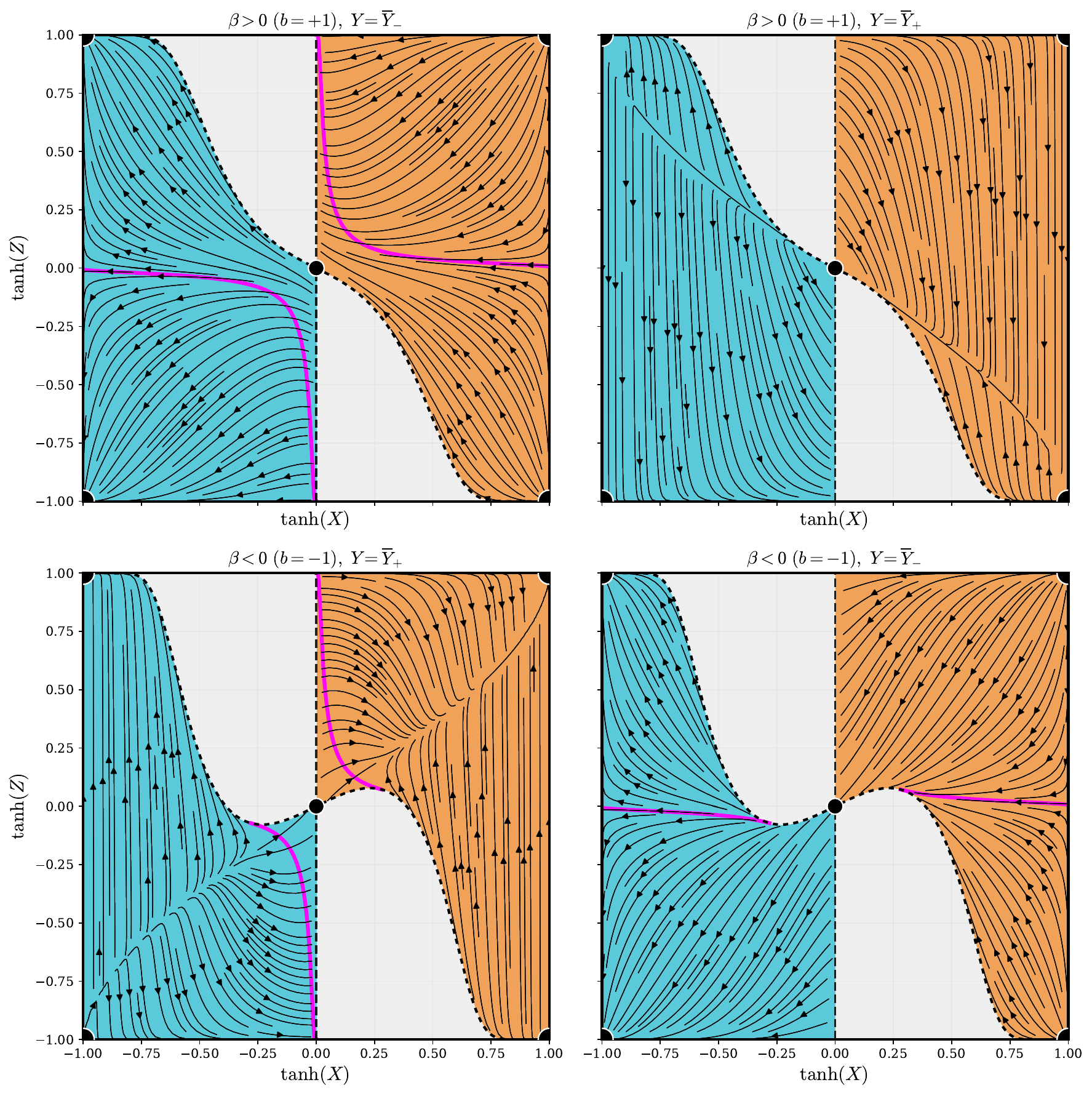}
\caption{Phase space flow corresponding to the dynamical system given 
in~(\ref{DynamicalSystem3}), with the top panels corresponding 
to~$\beta\!>\!0$, and the bottom panels corresponding to~$\beta\!<\!0$. 
The light blue and light orange colours signify disconnected sectors of 
the phase space, each with a definite sign of the Hubble rate,
corresponding to the equally coloured sectors in Fig.~\ref{PhaseSpace1}
and Fig.~\ref{PhaseSpace2}. The dashed black lines denote where the two 
top panels, and the two bottom panels are glued together, and the stream 
lines are seen to continue smoothly between panels. Magenta lines 
denote the singular surface~$Y\!=\!0$, which is crossed without 
obstructions by a considerable portion of the phase space flow lines 
from both sectors. The black dots denote fixed points.}
\label{PhaseSpace3}
\end{figure}

Taken together, these three projections provide complementary descriptions of
the same underlying phase-space surface. Their common message is that the
singular surface~$Y=0$ is not, in general, dynamically inaccessible. Rather,
regular cosmological trajectories can pass through it. This observation is
central for our purposes: although the singular surface is associated with a
degeneracy of the constraint structure, it can nevertheless be encountered in
dynamical evolution. The question of what becomes of perturbations in the
vicinity of such crossings is therefore unavoidable, and it is precisely this
question that motivates the present analysis.

\subsection{Constraint analysis for perturbations in cosmology}

The preceding subsection shows that the singular surface can be 
reached by regular FLRW background evolution in the Starobinsky model. 
This makes the role of perturbations especially important. Even if the 
homogeneous background evolves smoothly through the surface, the 
perturbative constraint structure may degenerate there. The question 
is therefore whether the regularity of the background evolution persists 
once inhomogeneous perturbations are included, or whether new singular 
behaviour appears in the perturbative sector.

A complete analysis of perturbations near such crossings is a broad 
problem that we leave for future work. Here we focus on the constraint 
structure of linearized perturbations around the FLRW backgrounds 
discussed above, and in particular on what happens as the background 
approaches the singular surface. The result is qualitatively different 
from the case studied in 
Sec.~\ref{sec: Perturbations around singular-surface solutions}, 
where the background lies on the singular surface for all time.

\medskip 

For the FLRW background in the Starobinsky model the canonical 
variables take the values
\begin{subequations}
\begin{align}
&
\overline{N} = 1 \, ,
\qquad
\overline{N}_i = 0 \, ,
\qquad
\overline{h}_{ij} = a^2 \delta_{ij} \, ,
\qquad
\overline{\mathcal{R}}_{ij} = 0 \, ,
\qquad
\overline{\mathcal{K}}_{ij} = - H \overline{h}_{ij} \, ,
\\
&
\frac{\overline{\pi}{}^{ij}}{\sqrt{\overline{h}}}
    =
    - \frac{\alpha \overline{R}{}^2}{6H} \, \overline{h}{}^{ij} \, ,
\qquad
\frac{\overline{\rho}{}^{ij}}{\sqrt{\overline{h}}} =
    - \frac{2 \overline{h}{}^{ij}}{\kappa^2}
	\big( 1 + 2 \alpha \kappa^2 \overline{R} \big) \, ,
	\qquad
	\overline{\lambda}_{ij} = 0
    \, ,
\label{FLRWbackground}
\end{align}
\end{subequations}
where the background 4-dimensional Ricci scalar 
is~$\overline{R}\!=\!6(\dot{H} \!+\! 2H^2)$.
In particular, the singular surface~(\ref{CriticalSurface}) is the 
surface on which~$\overline{\rho}=0$.

At the singular surface, the linearized Hamiltonian, momentum, and 
traceless constraints are
\begin{subequations}
\begin{align}
\mathcal{H}_{\scr (1)} \approx{}&
	\sqrt{\overline{h}} \, 
    \bigg[
    2 H \frac{\delta \pi}{\sqrt{\overline{h}}}
    -
    \frac{2 \delta \mathcal{K}}{3} 
    \frac{\overline{\pi}}{\sqrt{\overline{h}}} 
    +
    \frac{\delta h}{8\alpha\kappa^4}
    -
    \frac{1}{3} \Big( \overline{\nabla}{}^2
        \!- 
        \frac{1}{4\alpha\kappa^2}
        \Big)
    \frac{\delta \rho}{\sqrt{\overline{h}}}
    \bigg]
	\, ,
\\
\mathcal{H}^i_{\scr (1)} \approx{}&
	\frac{2}{3} \sqrt{\overline{h}} \, \overline{\nabla}{}^i
    \bigg[
    H \frac{\delta \rho}{\sqrt{\overline{h}}}
    -
    \frac{\delta \pi}{\sqrt{\overline{h}}}
    +
    \frac{\delta h}{6} \frac{ \overline{\pi} }{\sqrt{\overline{h}}}
    \bigg]
        \, ,
\\
\Phi^{ij}_{\scr (1)} \approx{}&
    \overline{H}{}^{ij}{}_{k\ell} \delta \rho^{k\ell}
    \, ,
\qquad \qquad
\Psi^{ij}_{\scr (1)} \approx
    \overline{H}{}^{ij}{}_{k\ell}
        \Big( \delta \pi^{k\ell} 
        + \frac{\overline{\pi}}{3} \delta h^{k\ell} \Big)
    \, .
\end{align}
\end{subequations}
Their equal-time Poisson brackets vanish weakly at the crossing, as 
shown in Table~\ref{SingularSurfaceBrackets}.
\begin{table}[h!]
\renewcommand{\arraystretch}{1.6}
\setlength\tabcolsep{0pt}
\centering
\begin{tabular}{ | w{c}{1.8cm} || w{c}{2.5cm} w{c}{2.5cm} w{c}{2.5cm} w{c}{2.5cm} |} 
\hline
	&
	$\mathcal{H}_{\scr (1)} \llbracket s \rrbracket $
	&
	$\mathcal{H}_{\scr (1)}^k \llbracket s_k \rrbracket$
	&
	$\Phi^{k\ell}_{\scr (1)} \llbracket s_{k\ell} \rrbracket$
	&
	$\Psi^{k\ell}_{\scr (1)} \llbracket s_{k\ell} \rrbracket$
\\
\hline\hline
	$\mathcal{H}_{\scr (1)} \llbracket q \rrbracket$
	&
	$0$
	&
	$0$
	&
	$0$
	&
	$0$
\\
	$\mathcal{H}_{\scr (1)}^i \llbracket q_i \rrbracket$
	&
	$0$
	&
	$0$
	&
	$0$
	&
	$0$
\\
	$\Phi^{ij}_{\scr (1)} \llbracket q_{ij} \rrbracket$
	&
	$0$
	&
	$0$
	&
	$0$
    &
    $0$
\\
	$\Psi^{ij}_{\scr (1)} \llbracket q_{ij} \rrbracket $
	&
	$0$
	&
	$0$
	&
	$0$
    &
    $0$
\\[0.8ex]
\hline
\end{tabular}
\caption{Poisson brackets among the linearized primary and secondary constraints in the Starobinsky model, evaluated at the singular surface~(\ref{CriticalSurface}). Each entry 
gives~$\tt \{ row , column \} \approx entry$.} \label{SingularSurfaceBrackets}
\end{table}

The vanishing of these equal-time brackets does not by itself imply 
that all constraints can be treated as ordinary first-class constraints 
at the crossing. The reason is that the linearized constraints depend 
explicitly on time through the background solution. Their consistency 
conditions therefore receive contributions not only from their Poisson 
brackets with the linearized Hamiltonian, but also from the explicit 
time dependence of the background.

Rather than repeating the full computation from the beginning, we use 
the exact evolution equation for the traceless secondary constraint, 
obtained from Eq.~(\ref{StarAlgebraSecondary}):
\begin{align}
\dot{\Psi}^{ij} \approx{}&
    \frac{ N H^{ijk\ell} }{6} \bigg[
    -
    2 \pi \mathcal{K}_{k\ell}
    +
    \rho \Big( \mathcal{R}_{k\ell} - 2  \mathcal{K}_{km} \mathcal{K}^m{}_\ell
            + \frac{2}{3} \mathcal{K}_{k\ell} \mathcal{K}
            - \frac{ \nabla_k \nabla_\ell N }{N} \Big)
    -
    \sqrt{h} \nabla_k \nabla_\ell \Big( \frac{\rho}{\sqrt{h}} \Big)
    \bigg]
\nonumber \\
&
    +
    \frac{ H^{ijk\ell} \rho}{6} \Big[
    2 \mathcal{K}_{k}{}^m \nabla_\ell N_m
    +
    N_m \nabla^m \mathcal{K}_{k\ell}
    \Big]
    -
    \frac{1}{6} \rho H^{ijk\ell} N \lambda_{k\ell}
    \, .
\label{StarPsiDot}
\end{align}
Expanding this equation to linear order around the FLRW background 
and then evaluating it at the singular surface~$\overline{\rho}=0$ gives
\begin{equation}
\dot{\Psi}_{\scr (1)}^{ij}
	\approx
    - \frac{1}{6} \Omega_{\scr (1)}^{ij}
    \, ,
\qquad
    \Omega_{\scr (1)}^{ij}
	\equiv
    \overline{H}{}^{ijk\ell}  
		\bigg[
		2\overline{\pi}
		\big( \delta\mathcal{K}_{k\ell}
        +
        H \delta h_{k\ell} \big)
	+
	\sqrt{\overline{h}} \, \overline{\nabla}_k \overline{\nabla}_\ell
		\Big( \frac{\delta\rho}{\sqrt{\overline{h}}} \Big)
		\bigg]
		\, .
\end{equation}
Away from the singular surface, the consistency condition
$\dot{\Psi}^{ij} \!\approx \!0$ determines the traceless Lagrange
multiplier~$\lambda_{ij}$. At the crossing this determination fails, 
because the coefficient multiplying~$\lambda_{ij}$ is proportional 
to~$\overline{\rho}$ and therefore vanishes.

One might therefore be tempted to 
interpret~$\Omega^{ij}_{\scr (1)} \!\approx\! 0$
as a tertiary constraint generated at the instant of crossing. This
interpretation is misleading. Unlike the constraints considered in
Sec.~\ref{sec: Perturbations around singular-surface solutions}, this 
condition would hold only at the instant at which the background 
trajectory intersects the singular surface. There is consequently no 
independent requirement that it be preserved as a constraint for all 
times. The crossing problem is therefore not described by an ordinary 
Dirac--Bergmann algorithm with a conserved set of constraints.

The same conclusion can be seen directly from the regular-region 
expression for the traceless Lagrange multiplier. Linearizing 
Eq.~(\ref{StarLambdaSolution}) around the FLRW background gives
\begin{align}
\delta \lambda_{ij} ={}&
    -
    \frac{\Omega^{\scr (1)}_{ij} }{ \overline{\rho} }
    +
    \overline{H}_{ij}{}^{k\ell} \bigg[
    2H \big( \delta \mathcal{K}_{k\ell} + H \delta h_{k\ell} \big)
    +
    \frac{1}{2} \Big(
        2 \overline{\nabla}_k \overline{\nabla}{}^m \delta h_{\ell m}
        - \overline{\nabla}_k \overline{\nabla}_{\ell} \delta h
        - \overline{\nabla}{}^2 \delta h_{k\ell}
        \Big)
\nonumber \\
&
\hspace{3.5cm}
    +
    \frac{1}{6} \overline{\nabla}_k \overline{\nabla}_\ell \delta h
        -
        \overline{\nabla}_k \overline{\nabla}_\ell \delta N
    -
    2 H \overline{\nabla}_k \delta N_\ell
    \bigg]
    \, .
\end{align}
Thus, regular propagation through the crossing requires
$\Omega^{ij}_{\scr(1)}/\overline{\rho}$ to remain finite as
$\overline{\rho}\!\to\!0$. Equivalently, the numerator
$\Omega^{ij}_{\scr(1)}$ must approach zero at least as fast as the
background quantity~$\overline{\rho}$. This is not a standard
constraint-preservation condition, but a regularity condition at a singular
point of the perturbative evolution equations.

This is qualitatively different from the situation analyzed in
Sec.~\ref{sec: Perturbations around singular-surface solutions} and 
in Ref.~\cite{Barker:2025gon}, where the background lies on the singular 
surface for all time. There the perturbative constraint structure 
differs from the one in the regular region of phase space, but it does 
not change along the background trajectory. A standard classification 
into first- and second-class constraints is therefore possible. In the 
present case, by contrast, the rank of the linearized constraint algebra 
changes only at the crossing. The usual Dirac--Bergmann algorithm does 
not by itself provide a stable constraint classification through that 
instant, and consequently there is no ready-made way to assign a 
standard number of propagating degrees of freedom exactly at the 
singular surface.

\section{Discussion}
\label{sec: Discussion}

We first considered the Hamiltonian constraint analysis of
general~$f(R)$ theories in the Jordan frame, performed in
Sec.~\ref{sec: Hamiltonian constraint analysis}. One of the main
results of this work is the identification of singular surfaces in
phase space where the constraint structure changes discontinuously.
These singular surfaces are located at~$f'(R) \!=\! 0$ and~$f''(R) \!=\! 0$,
where second-class sectors of the theory change character to
first-class. This analysis extends and complements the corresponding
analysis of pure~$R^2$ gravity~\cite{Barker:2025gon}, showing that
the singular behaviour previously found in that model is not an
isolated peculiarity, but part of a broader phase-space structure
present in~$f(R)$ theories.

It is important to stress that this is an intrinsic Jordan-frame
result. Although the same loci coincide with singular points of the
field redefinition relating the Jordan and Einstein frames, the
degeneracy of the constraint structure is not caused by that
transformation. Rather, it is a property of the Jordan-frame
canonical theory itself. Related behaviour is also expected in
metric-affine~$f(R)$ theories, as already indicated for the
pure~$R^2$ model in~\cite{Karananas:2024qrz}. In that setting, the
same phenomenon can be interpreted as the appearance of a
phase-space singular surface in the constraint analysis
of~\cite{Glavan:2023cuy}.

We then examined the behaviour of linear perturbations in the
vicinity of the identified singular surfaces. In
Sec.~\ref{sec: Perturbations around singular-surface solutions},
we considered~$f(R)$ models that admit solutions
with~$f'(R) \!=\! 0$, and studied perturbations around such
backgrounds. We found that these models exhibit an empty spectrum
of linearized perturbations around such backgrounds. This was
already noted for maximally symmetric backgrounds in such models
in~\cite{Casado-Turrion:2023rni,Casado-Turrion:2024esi}.
Here we extend this result to arbitrary backgrounds of this
type\footnote{Note that the constant Ricci scalar does not
imply that the solution has to be maximally symmetric (anti-)de Sitter
or Minkowski spacetime; e.g. Schwarzschild-de Sitter spacetime has
a constant Ricci scalar.} by deriving and analyzing the Hamiltonian
constraint structure in
Sec.~\ref{sec: Perturbations around singular-surface solutions}.
The linearized spectrum is empty because the traceless pair of
constraints changes character from second-class to first-class,
while the Hamiltonian and momentum constraints degenerate into only
two independent first-class constraints.
Furthermore, if the background in addition satisfies~$f''(R)\!=\!0$
then the linearized scalar sector degenerates differently --- the
primary scalar constraint becomes first-class as the secondary scalar 
constraint ceases to provide an independent constraint. Despite
these different constraint structures, the resulting perturbative
spectrum is empty in both cases.

This analysis further establishes that the behaviour of perturbations found
in pure~$R^2$
theory~\cite{Hell:2023mph,Golovnev:2023zen,Karananas:2024hoh,Barker:2025gon}
is not a special property of a particular background in that theory.
Rather, it is a general feature of a large class of~$f(R)$ models
that allow for solutions satisfying~$f'(R) \!=\! 0$. This class
includes the pure~$R^2$ model as a special case, while the
Starobinsky model illustrates how the same singular surface persists
after adding an Einstein-Hilbert term. In that case the singular
surface does not disappear, but is instead shifted to a nonvanishing
value of the Ricci scalar in~(\ref{CriticalSurface}). In the
limit in which the Starobinsky model reduces to the pure~$R^2$
model, the singular surface moves smoothly to~$R \!=\! 0$.

Finally, in
Sec.~\ref{sec: Perturbations near singular-surface crossings}, we
examined the Starobinsky model, which admits backgrounds that
approach a singular surface dynamically, and studied what this
implies for perturbations propagating on such backgrounds. This
question is substantially more subtle than the case of backgrounds
that lie entirely on a singular surface. One must first identify
backgrounds that actually evolve through a singular surface, and
then determine how perturbations behave as the surface is
approached. In this setting, even the split into background and
perturbations becomes delicate. Here we addressed this intricate
question only in the restricted setting of FLRW backgrounds in the
Starobinsky model. By analyzing the phase-space diagrams in
Sec.~\ref{sec: Cosmological phase space analysis}, we showed that
this background admits solutions that cross the singular
surface~$R \!=\! - 1/(2\alpha\kappa^2)$.

We then examined how perturbations behave on such a cosmological
background as it approaches the singular surface. The constraint
structure of linearized perturbations changes discontinuously, but
in a manner very different from the one described in
Sec.~\ref{sec: Perturbations around singular-surface solutions}.
At the moment of the singular-surface crossing, the bracket between
the two traceless second-class constraints vanishes. Consequently,
the traceless Lagrange multiplier is no longer determined in the
standard way by the conservation of the secondary traceless
constraint. However, this does not mean that the two traceless
constraints can simply be reclassified as first-class. Nor does it
mean that the conservation of the secondary traceless constraint
generates an ordinary tertiary constraint. Instead, the rank of the
constraint algebra changes along the background trajectory, and the
standard Dirac--Bergmann algorithm does not by itself provide a
stable degree-of-freedom count at the crossing. For this reason, we
found no sensible way to assign a standard number of propagating
degrees of freedom exactly at the singular surface.

What does emerge from the analysis is a necessary condition for the
regular propagation of perturbations through the crossing. A certain
quantity must vanish at the singular surface, and it must approach
zero at least as fast as~$f'(R)$ approaches zero. This condition
does not fit naturally into the usual Dirac--Bergmann classification.
It is instead more reminiscent of regularity conditions at singular
points of differential equations, where solutions must take special
forms in order to pass smoothly through the singular point. It
therefore remains unclear whether perturbations can in fact
propagate through the crossing when the background reaches the
singular surface.

Thus, dynamically approaching a singular surface is qualitatively
different from perturbing around an exact solution that lies on the
singular surface for all time. This points to the main question
stemming from this analysis: what is the behaviour of perturbations
as they approach the singular surface? Do they shield the dynamics
from ever approaching singular surfaces, or do several effects
conspire to allow the evolution to pass through the crossing?
Lessons from studies of other modified gravity theories with
singular surfaces such as Einsteinian Cubic
Gravity~\cite{BeltranJimenez:2020lee} suggest that the former is
definitely possible. However, answering this question for~$f(R)$
theories is left for future work. What we have established here is
that this question is pertinent to a broad class of~$f(R)$ models,
and that a perturbative degree-of-freedom count tied to a particular
background can fail to capture the underlying canonical structure of
the theory.

\section*{Acknowledgments}
\addcontentsline{toc}{section}{\protect\numberline{}Acknowledgments} 

DG was supported by the Czech Science Foundation (GA\v{C}R) grant 24-13079S.
This work was co-financed by the European Structural and Investment Funds and the Czech Ministry of Education, Youth and Sports (MSMT) of the Czech Republic 
 (Project FORTE – $\mathrm{PCZ.02.01.01/00/22\_008/0004632}$).


\end{document}